%% file: JRNL_BOSSAMP.tex
\documentclass[conference]{IEEEtran}

\usepackage{cite}
\usepackage[dvips]{graphicx}

\usepackage[cmex10]{amsmath}
\usepackage{amsfonts}
\usepackage{bm}

\usepackage{pifont}
\newcommand{\cmark}{\ding{51}}
\newcommand{\xmark}{\ding{55}}

\usepackage{array}
\usepackage[printonlyused]{acronym} 

\usepackage{algorithm}
\usepackage{algorithmicx}
\usepackage[noend]{algpseudocode}
\algdef{SE}[DOWHILE]{Do}{doWhile}{\algorithmicdo}[1]{\algorithmicwhile\ #1}%

\usepackage{enumitem}
\usepackage{xcolor}

\input{acronymes.tex}
\input{definitions.tex}

\usepackage[bookmarks=false]{hyperref}
\hypersetup{
  pdffitwindow=       true,
  pdfcenterwindow=    true,
  pdfstartview=       FitH,
  bookmarksopen=      true,   
  bookmarksnumbered=  true,  
  pdfhighlight=       /P,
  linktocpage=        true,   
  colorlinks=         false,
  pdftitle=           {Bayesian Optimal Approximate Message Passing to Recover Structured Sparse Signals},
  pdfsubject=         {Vienna University of Technology},
  pdfauthor=          {Martin Mayer, Norbert Goertz},
  pdfkeywords=        {compressed sensing, message passing, joint sparse, group sparse, binary, Gaussian mixtures, extrinsic update},
  pdfborder=          0 0 0,
  linkcolor=          xBlue,
  citecolor=          xBlue,
  urlcolor=           xBlue
  }

\usepackage{cleveref}
\crefname{appsec}{Appendix}{Appendices}

\crefformat{chapter}{#2Chapter~#1#3}
\crefformat{section}{#2Section~#1#3}
\crefformat{subsection}{#2Section~#1#3}
\crefformat{Appendix}{#2Appendix~#1#3}
\crefformat{figure}{#2Figure~#1#3}
\crefformat{table}{#2Table~#1#3}
\crefformat{equation}{#2(#1)#3}
\crefmultiformat{equation}{#2(#1)#3}{ and~#2(#1)#3}{, ~#2(#1)#3}{, and~#2(#1)#3}
\crefmultiformat{figure}{#2Figures~#1#3}{ and~#2#1#3}{, ~#2#1#3}{, and~#2#1#3}
\crefmultiformat{table}{#2Tables~(#1)#3}{ and~#2(#1)#3}{, ~#2(#1)#3}{, and~#2(#1)#3}
\crefrangeformat{equation}{#3(#1)#4--#5(#2)#6}
\crefrangeformat{figure}{#3Figures~#1#4--#5#2#6}

\begin{document}
\title{Bayesian Optimal Approximate Message Passing to Recover Structured Sparse Signals}

\author{\IEEEauthorblockN{Martin Mayer, \emph{Student Member}, \emph{IEEE}, and Norbert Goertz, \emph{Senior Member}, \emph{IEEE}}}

\maketitle

\begin{abstract}
	\input{abstract.tex}
\end{abstract}

\begin{IEEEkeywords}
compressed sensing, message passing, group sparse, jointly sparse, sparse binary, Bernoulli-Gaussian, Gaussian mixture, extrinsic information, turbo decoding
\end{IEEEkeywords}

\acresetall

\IEEEpeerreviewmaketitle

\section{Introduction}
 \label{sec:introduction}
 \input{introduction.tex}

\section{Recovery of Sparse Signals}
\label{sec:recovery_sparse}
\input{recovery_sparse.tex}

\section{Recovery of Group Sparse Signals}
\label{sec:recovery_sparse_group}
\input{recovery_sparse_group.tex}

\section{Recovery of Jointly Sparse Signals}
\label{sec:recovery_sparse_joint}
\input{recovery_sparse_joint.tex}

\section{Recovery of Arbitrary Structured Signals}
\label{sec:recovery_general}
\input{recovery_general.tex}

\section{Comparison and Numerical Results}
\label{sec:results}
\input{results.tex}

\section{Conclusion}
\label{sec:conclusion}
\input{conclusion.tex}


\bibliographystyle{ieeetr}
\bibliography{JRNL_BOSSAMP}

\end{document}

%% file: acronymes.tex
\acrodef{AMP}[AMP]{Approximate Message Passing}
\acrodef{BAMP}[BAMP]{Bayesian optimal Approximate Message Passing}
\acrodef{BOSSAMP}[BOSSAMP]{Bayesian Optimal Structured Signal Approximate Message Passing}
\acrodef{PDF}[PDF]{Probability Density Function}
\acrodef{PMF}[PMF]{Probability Mass Function}
\acrodef{ADMM}[ADMM]{Alternating Direction Method of Multipliers}
\acrodef{GAMP}[GAMP]{Generalized Approximate Message Passing}
\acrodef{HGAMP}[HGAMP]{Hybrid Generalized Approximate Message Passing}

\acrodef{LLR}[LLR]{Log Likelihood Ratio}

\acrodef{EM}[EM]{Expectation Maximization}

\acrodef{MAP}[MAP]{Maximum Aposteriori Probability}
\acrodef{MMSE}[MMSE]{Minimum Mean Squared Error}

\acrodef{NMSE}[NMSE]{Normalized Mean Squared Error}
\acrodef{FANMSE}[FANMSE]{False Alarm NMSE}

\acrodef{RIP}[RIP]{Restricted Isometry Property}
\acrodef{SNR}[SNR]{Signal-to-Noise Ratio}
\acrodef{MSE}[MSE]{Mean Squared Error}
\acrodef{LASSO}[LASSO]{Least Absolute Shrinkage and Selection Operator}
\acrodef{CRC}[CRC]{Cyclic Redundancy Check}

%% file: definitions.tex
\newcommand{\gma}{\gamma} 
\newcommand{\gmavec}{\boldsymbol{\gma}} 
\newcommand{\gmamat}{\boldsymbol{\Gamma}} 

\newcommand{\resi}{\mathbf{r}} 

\newcommand{\varc}{\beta} 

\newcommand{\xscalar}{x}
\newcommand{\yscalar}{y}
\newcommand{\wscalar}{w}

\newcommand{\Xscalar}{\mathsf{x}}
\newcommand{\Yscalar}{\mathsf{y}}
\newcommand{\Wscalar}{\mathsf{w}}
\newcommand{\Uscalar}{\mathsf{u}}

\newcommand{\xvec}{\mathbf{\xscalar}}
\newcommand{\yvec}{\mathbf{\yscalar}}
\newcommand{\wvec}{\mathbf{\wscalar}}

\newcommand{\xmat}{\mathbf{X}}
\newcommand{\ymat}{\mathbf{Y}}
\newcommand{\wmat}{\mathbf{W}}

\newcommand{\Xvec}{\boldsymbol{\Xscalar}}
\newcommand{\Yvec}{\boldsymbol{\Yscalar}}
\newcommand{\Wvec}{\boldsymbol{\Wscalar}}

\newcommand{\xdim}{N}
\newcommand{\ydim}{M}

\newcommand{\xind}{n} 
\newcommand{\yind}{m} 

\newcommand{\pdf}{f}
\newcommand{\pmf}{p}
\newcommand{\jointpdf}{\pdf_{\Xvec,\Yvec}}
\newcommand{\priorpdf}{\pdf_{\Xvec}}
\newcommand{\posteriorpdf}{\pdf_{\Xvec|\Yvec}}
\newcommand{\likelihoodpdf}{\pdf_{\Yvec|\Xvec}}
\newcommand{\ypdf}{\pdf_{\Yvec}}

\newcommand{\A}{\mathbf{A}}

\newcommand{\varx}{\sigma_{\Xscalar_\xind}^2}
\newcommand{\varxextr}{\sigma_{\Xscalar_{l}}^2}

\newcommand{\varxl}{\sigma_{\Xscalar_{\xind,l}}^2}

\newcommand{\numgroups}{{N_G}} 
\newcommand{\numblocks}{{N_B}} 

\newcommand{\support}{\mathcal{S}}
\newcommand{\compsupport}{\overline{\mathcal{S}}}
\newcommand{\group}{\mathcal{G}}

\newcommand{\edges}{\mathcal{E}}
\newcommand{\variables}{\mathcal{V}}
\newcommand{\factors}{\mathcal{F}}
\newcommand{\factorgraph}{\text{FG}}

\newcommand{\groupupdate}{U_G}

\newcommand{\priorupdate}{U_P}

\newcommand{\latent}{z} 
\newcommand{\Latent}{\mathsf{z}} 

\newcommand{\transp}{{\mathrm{T}}}

\newcommand{\NMSE}{\text{NMSE}}
\newcommand{\FANMSE}{\text{FANMSE}}

%% file: abstract.tex
We present a novel compressed sensing recovery algorithm -- termed \ac{BOSSAMP} -- that jointly exploits the prior distribution and the structured sparsity of a signal that shall be recovered from noisy linear measurements.
Structured sparsity is inherent to \emph{group sparse} and \emph{jointly sparse} signals.
Our algorithm is based on approximate message passing that poses a low complexity recovery algorithm whose Bayesian optimal version allows to specify a prior distribution for each signal component.
We utilize this feature in order to establish an iteration-wise \emph{extrinsic group update} step, in which likelihood ratios of neighboring group elements provide soft information about a specific group element.
Doing so, the recovery of structured signals is drastically improved.

We derive the extrinsic group update step for a \emph{sparse binary} and a \emph{sparse Gaussian} signal prior, where the nonzero entries are either one or Gaussian distributed, respectively. 
We also explain how \ac{BOSSAMP} is applicable to arbitrary sparse signals.

Simulations demonstrate that our approach exhibits superior performance compared to the current state of the art, while it retains a simple iterative implementation with low computational complexity.

%% file: introduction.tex
Solving a linear system of equations $\yvec = \mathbf{A}\xvec$ for $\xvec$ is an omnipresent problem in various fields, as a myriad of problem statements can be written in such form.
While the very classical techniques such as least squares or \ac{MMSE} estimation minimize the error with respect to the $\ell_2$-norm, the incorporation of additional knowledge like sparsity and structure of $\xvec$ has gained a lot of attention over recent years.

In particular, compressed sensing was introduced in \cite{donoho2006compressed,candes2006robust,candes2006stable} to solve a linear system of equations in case of a sparsely populated $\xvec$, i.e., the vector features only few nonzero entries. 
A change of paradigm was triggered by recognizing that a sparse $\xvec$ can be reconstructed perfectly from an \emph{underdetermined} system of linear equations.
In the context of signal processing, a signal vector $\xvec$ can thus be reconstructed from undersampling, where the sampling basis functions are the columns of $\mathbf{A}$, and where the samples are stored in $\yvec$.
This led to a huge popularity in the search for efficient recovery algorithms that incorporate the sparsity constraint; the classical compressed sensing formulation 
\begin{equation*}
\widehat{\xvec} = \arg \min_{\widetilde{\xvec} \in \mathbb{R}^\xdim} \left\| \widetilde{\xvec} \right\|_0 \ \ \text{s.t.} \ \ \mathbf{A}\widetilde{\xvec} = \yvec
,
\end{equation*}
where the pseudo norm $\left\| \widetilde{\xvec} \right\|_0$ counts the number of nonzero entries in $\widetilde{\xvec}$, leads to a combinatorial search for $\widehat{\xvec}$ and is generally NP-hard to solve.
The problem was relaxed into an $\ell_1$-norm minimization called basis pursuit (denoising) \cite{chen1998atomic} which is also applicable to noisy samples $\yvec$.
An alternative formulation that introduces a controllable weight for the sparsity constraint is given by the \ac{LASSO} \cite{tibshirani1996regression}.
A computationally efficient recovery algorithm that iteratively solves the \ac{LASSO} is provided by \ac{AMP} that was introduced in \cite{donoho2009message,maleki2010approximate,donoho2010message,donoho2011design}. 
Its foundation is Gaussian loopy belief propagation \cite{weiss2001correctness} with simplified message passing that assumes high dimensional signal vectors $\xvec$.
Its Bayesian optimal version that exploits the signal prior is described in \cite{maleki2010approximate,donoho2010message,donoho2011design}, we will henceforth denote it as \ac{BAMP}.
\ac{BAMP} was further extended in \cite{rangan2011generalized} in order to allow for non-linear and non-Gaussian output relations; the samples $\yvec$ are transformed by an arbitrary but known function, and the output of the transformation is known to the estimator.

In many problems, $\xvec$ features a certain structure in the sparsity, i.e., collections of entries (groups) contain either only zeros or only nonzero entries. 
Group sparsity typically occurs in (image) classification tasks \cite{zhang2010automatic,ng2011generalized,gui2014group}.
To account for this, the \ac{LASSO} was extended to the group \ac{LASSO} \cite{yuan2006model,friedman2010note,boyd2011distributed,deng2013group}.
Another established scheme is grouped orthogonal matching pursuit \cite{swirszcz2009grouped}.
In the realm of \ac{AMP}, the generalized approach \cite{rangan2011generalized} was extended in \cite{rangan2012hybrid} to incorporate the signal structure.

Aside from group sparsity, structure is also found in jointly sparse signals, i.e., $\numblocks$ vectors $\xvec_b$ share a common support (the nonzero entries occur at the same indices, $\forall b\in\{1,...,\numblocks\}$). 
A prominent instance of this is the multiple measurement vector problem \cite{cotter2005sparse,ziniel2013efficient}.
Typical applications of the jointly sparse case are neuromagnetic imaging \cite{cotter2005sparse,liang2009parallel} and direction-of-arrival estimation \cite{tzagkarakis2010multiple}.

\subsection{Contributions}

In this paper, we present a novel recovery algorithm -- termed \ac{BOSSAMP} -- that extends \ac{BAMP} to incorporate the signal structure, such as group or joint sparsity.
The approach also allows for overlapping groups, and it is not restricted to the sparse case.
The key feature is the inclusion of a \emph{group update} step that is inspired by an \emph{extrinsic information} exchange that is predominantly used in coding \cite{hagenauer1995source,hagenauer1996iterative,hagenauer2004exit,berrou1996near}, where it is also known as the \emph{turbo principle}. 
In each iteration of \ac{BOSSAMP}, the probability that a specific group entry was zero is updated by accumulating the extrinsic information of all other group entries in terms of likelihood ratios.
This leads to a superior recovery performance compared to other state of the art approaches such as \cite{swirszcz2009grouped, rangan2012hybrid ,deng2013group, boyd2011distributed}, which we show by simulation.
Furthermore, our algorithm converges to a solution in very few iterations, while its implementation stays simple and efficient. 
Specifically, it only requires two matrix-vector multiplications per iteration, matrix inversions are not required.

\subsection{Related Work and Novelty}

Merging (loopy) belief propagation with turbo equalization to recover structured sparse signals has already been suggested in \cite{schniter2010turbo}, where the factor graph of the observation structure was extended by a pattern structure. 
Hidden binary indicators were used to model whether signal entries are active (nonzero) or inactive (zero).
Sparsity pattern beliefs are exchanged between the observation structure and the sparsity structure in an iterative manner by leveraging (loopy) belief propagation. 
This approach was later utilized in compressive imaging \cite{som2012compressive} to exploit the sparsity and persistence accross scales of 2D wavelet coefficients of natural images.
A generalized manifestation of \ac{AMP} that embeds the ideas of \cite{schniter2010turbo} and presents an algorithmic implementation is provided in \cite{rangan2012hybrid}, to which we compare our scheme to.

While \cite{schniter2010turbo,som2012compressive} approach the topic from the message passing point of view, we focus on an alternative description that utilizes likelihood ratios in terms of $L$-values.
We utilize the classic \ac{BAMP} algorithm and extend the iteration loop by two steps, namely the \emph{group update} and the subsequent \emph{prior update}.
On the one hand, \ac{BAMP} assumes independently but non-identically distributed signal entries in $\mathbf{x}$ to perform scalar \ac{MMSE} estimation. 
On the other hand, our two additional steps update the entry-wise prior information for the next \ac{BAMP} iteration by exploiting the structure in the sparsity. 
In a ''ping-pong'' manner, an \ac{MMSE} estimate emerges.

We give a detailed and easy-to-follow derivation of the group update step, specifically for the following two prominent cases, where we provide simple closed form expressions:
first for the \emph{sparse binary} case, where $L$-values can be formulated naturally, and then
for the \emph{sparse Gaussian} (also known as Bernoulli-Gaussian) case, where we introduce a latent binary variable that indicates whether an entry was zero or nonzero. 

While the resulting \ac{BOSSAMP} algorithm shares similarities with the \ac{HGAMP} algorithm \cite{rangan2012hybrid} which is applicable to a more general class of problems, our approach sticks to the standard \ac{AMP} framework \cite{donoho2009message,maleki2010approximate,donoho2010message,donoho2011design} which is mainly applicable to samples that are corrupted by Gaussian noise. 
While being restricted to a smaller class of problems, this leads to a \emph{simpler implementation} and a \emph{higher comprehensibility}.
Furthermore, simulation results suggest that our approach outperforms \ac{HGAMP} in terms of recovery performance and phase transitions.

\subsection{Notation}

Boldface letters such as $\mathbf{A}$ and $\mathbf{a}$ denote matrices and vectors, respectively.
Considering matrix $\mathbf{A}$, $\mathbf{A}_{i,:}$ is its $i$-th row, while $\mathbf{A}_{:,j}$ is its $j$-th column.
The superscript $(\cdot)^\transp$ denotes the transposition of a matrix or vector.
The vectorization of an $M\times N$ matrix is denoted $\mathbf{A}(:) \equiv [\mathbf{A}_{:,1}^\transp,...,\mathbf{A}_{:,N}^\transp]^\transp$.
The \mbox{$N \times N$} identity matrix is denoted $\mathbf{I}_N$.
The length $N$ all-one vector is denoted $\mathbf{1}_N$, while the $N\times N$ all-one matrix is denoted $\mathbf{1}_{N\times N}$.
Similarly, we define the all-zero vector $\mathbf{0}_N$ and the all-zero matrix $\mathbf{0}_{N\times N}$.
Calligraphic letters $\mathcal{S}$ denote sets,
their usage as subscript $\mathbf{a}_{\mathcal{S}}$ implies that only the vector entries defined by the elements in $\mathcal{S}$ are selected.
The cardinality of a set is denoted by \mbox{$|\mathcal{S}|$}.
Random variables and vectors are denoted by sans serif font as $\mathsf{x}$ and $\boldsymbol{\mathsf{x}}$, respectively.
While \mbox{$\Xscalar \sim \mathcal{N}(\mu,\sigma^2)$} denotes a Gaussian distributed random variable $\Xscalar$ with mean $\mu$ and variance $\sigma^2$, the shorthand notation
\begin{equation}
\mathcal{N}(x|\mu,\sigma^2) \equiv \frac{1}{\sqrt{2\pi\sigma^2}}\exp{\left(-\frac{1}{2\sigma^2}(x-\mu)^2\right)}
\end{equation}
denotes that such a Gaussian distribution is evaluated at the value $x$.

\subsection{Outline}

The remainder of this paper is outlined as follows:
\cref{sec:recovery_sparse} reviews compressed sensing and draws the link between the probabilistic graphical model of the estimation and the iterative recovery schemes \ac{AMP} and \ac{BAMP}, the algorithmic implementation of these is also presented. 
Finally, two prominent sparse signal priors, namely the sparse binary and the sparse Gaussian prior, are introduced.
\cref{sec:recovery_sparse_group} describes the group sparse case and presents the corresponding \ac{BOSSAMP} algorithm, whose update rules are derived for the sparse binary and the sparse Gaussian signal prior, respectively.
\cref{sec:recovery_sparse_joint} does the same for the jointly sparse case.
\cref{sec:recovery_general} discusses the extension of \ac{BOSSAMP} to the arbitrary signal case.
\cref{sec:results} introduces the figures of merit and the comparative schemes for simulation, and then presents and discusses the numerical results.
The paper is concluded in \cref{sec:conclusion}.

%% file: recovery_sparse.tex
\subsection{Compressed Sensing}

Compressed sensing was introduced in \cite{donoho2006compressed,candes2006robust,candes2006stable} to reconstruct a high-dimensional signal vector 
\mbox{$\mathbf{x} \in \mathbb{R}^{N}$}
from \mbox{$M<N$} linear measurements
\begin{equation}
\mathbf{y} = \mathbf{A}\mathbf{x} + \mathbf{w}
,
\label{eq:cs_measurement}
\end{equation}
where 
\mbox{$\mathbf{y} \in \mathbb{R}^{M}$} 
is the measurement vector,
\mbox{$\mathbf{A} \in \mathbb{R}^{M \times N}$} 
is the fixed sensing~matrix and
\mbox{$\mathbf{w} \in \mathbb{R}^{M}$}
is additive measurement noise.
Signal vector $\mathbf{x}$ is assumed to be $K$-sparse: it has at most $K$ out of $N$ nonzero entries, where \mbox{$K \ll N$}.
This enables to recover $\mathbf{x}$ from \cref{eq:cs_measurement} although the system of equations is underdetermined.
To ensure stable recovery from noisy measurements, the sensing~matrix~$\mathbf{A}$ has to satisfy the \ac{RIP} \cite{candes2006stable} 
\begin{equation}
(1-\delta_{2K}) \left\| \mathbf{v} \right\|_2^2 \leq \left\| \mathbf{A}\mathbf{v} \right\|_2^2 \leq  (1+\delta_{2K}) \left\| \mathbf{v} \right\|_2^2
\label{eq:RIP}
\end{equation}
of order $2K$ and level \mbox{$\delta_{2K}\in(0,1)$}
for all $2K$-sparse vectors $\mathbf{v}$, which basically implies that the linear operator $\mathbf{A}$ preserves the Euclidean distance between every pair of $K$-sparse vectors up to a small constant $\delta_{2K}$.

An appropriate sensing matrix can be constructed by picking $\mathbf{A}$ randomly with i.i.d. \mbox{(sub-)}Gaussian entries.
Such matrices were proven in \cite{baraniuk2008simple} to almost surely satisfy the \ac{RIP}
while the number of required measurements for successful recovery is lower bounded by
\begin{equation}
	M = \left\lceil c K \log \frac{N}{K} \right\rceil
	,
\label{eq:required_measurements}
\end{equation}
where $c$ is a small constant and the ceiling operation $\lceil \cdot \rceil$ ensures an integer number of measurements.

\subsection{Graphical Model}

A graphical model poses the probabilistic foundation of compressed sensing recovery algorithms that are based on message passing, such as \ac{AMP} and \ac{BAMP}, see \cite{montanari2012graphical}.
The key ingredient is a factorization of a multivariate distribution -- in our case the posterior distribution of $\xvec$ given $\yvec$ based on \cref{eq:cs_measurement} -- into many factors.
This factorization is described and visualized by a factor graph \cite{yedidia2011message, loeliger2007factor}, which we will now present.

Let us begin with the underlying probabilistic assumptions.
Considering \cref{eq:cs_measurement}, we assume that only sensing matrix $\mathbf{A}$ is deterministic (fixed), which leaves us to characterize the distributions of measurement vector $\mathbf{y}=[y_1,...,y_m,...,y_M]^\transp$, 
signal vector $\mathbf{x}=[x_1,...,x_n,...,x_N]^\transp$ 
and noise vector $\mathbf{w}=[w_1,...,w_m,...,w_M]^\transp$.

We assume white Gaussian noise with zero mean and covariance $\sigma_\Wscalar^2\mathbf{I}_M$, i.e., \mbox{$\Wvec \sim \mathcal{N}(\mathbf{0}, \sigma_\Wscalar^2\mathbf{I}_M)$}. The noise \ac{PDF} calculates as
\begin{equation}
\pdf_{\Wvec}(\wvec) 
=  
\prod_{\yind=1}^{\ydim} \pdf_{\Wscalar}(\wscalar_\yind)
=
\prod_{\yind=1}^{\ydim} 
\mathcal{N}(\wscalar_\yind|0,\sigma_\Wscalar^2)
.
\label{eq:noise_pdf}
\end{equation}
The joint \ac{PDF} of signal and measurement can be factored according to Bayes' rule:
\begin{equation}
\jointpdf(\xvec,\yvec) 
= \underbrace{\posteriorpdf(\xvec|\yvec)}_{\text{posterior}} \ypdf(\yvec)
= \underbrace{\likelihoodpdf(\yvec|\xvec)}_{\text{likelihood}} \underbrace{\priorpdf(\xvec)}_{\text{prior}}
.
\label{eq:joint_pdf}
\end{equation}
We assume independently distributed signal entries with \ac{PDF} 
$\pdf_{\Xscalar_\xind}(\xscalar_\xind)$, the \emph{prior}, thus, factors as
\begin{equation}
\priorpdf(\xvec) = \prod_{\xind=1}^{\xdim} \pdf_{\Xscalar_\xind}(\xscalar_\xind)
.
\label{eq:prior}
\end{equation}
The \emph{likelihood} is characterized by the noise \ac{PDF}:
\begin{equation}
\begin{aligned}
\likelihoodpdf(\yvec|\xvec) 
= \pdf_{\Wvec}(\yvec-\A\xvec) 
=  \prod_{\yind=1}^{\ydim} \pdf_{\Wscalar}\left(\yscalar_\yind-\A_{\yind,:}\xvec\right) 
.
\label{eq:likelihood}
\end{aligned}
\end{equation}
Estimators of $\xvec$ typically rely on the \emph{posterior} 
\begin{equation}
\posteriorpdf(\xvec|\yvec) = 
\prod_{\yind=1}^{\ydim}
\frac{1}{\pdf_{\Yscalar_\yind}(\yscalar_\yind)}
\pdf_{\Yscalar_\yind|\Xvec}(\yscalar_\yind|\xvec)
\pdf_{\Xvec}(\xvec)
\label{eq:posterior_expanded}
\end{equation}
that entails $M$ factors.

\begin{figure}[t]
\includegraphics[width=0.49\textwidth]{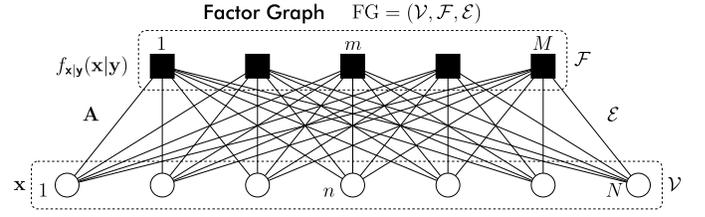} 
\caption{Factor graph of measurement \cref{eq:cs_measurement} associated to \cref{eq:posterior_expanded}.}
\label{fig:factor_graph}
\end{figure}

The resulting factor graph $\factorgraph=(\variables,\factors,\edges)$ consists of the 
variable nodes $\variables=\{1,...,\xdim\}$ that encompass the signal of interest $\xvec$,
the factor nodes $\factors=\{1,...,\ydim\}$ associated to \cref{eq:posterior_expanded}, 
and the edges $\edges=\{ (\yind,\xind) : \yind\in \factors, \xind \in \variables \}$ that 
correspond to the relations between $\xvec$ and $\yvec$ which are dictated by the entries of $\A$, i.e., $\A_{\yind,\xind}$.
Since typically, all entries of $\A$ are nonzero, our bipartite graph, as illustrated by \cref{fig:factor_graph}, is fully connected.

As we intend to recover $\xvec$ from $\yvec$ knowing $\A$, 
we formulate the \ac{MMSE} estimator
\begin{equation}
\widehat{\xvec}_{\text{MMSE}}(\yvec)
=\mathbb{E}_{\Xvec}\left[ \Xvec|\Yvec = \yvec \right] 
=\int_{\mathbb{R}^{\xdim}}  \!\! \widetilde{\xvec} \,  \posteriorpdf(\widetilde{\xvec}|\yvec) \, d\widetilde{\xvec} \\
.
\label{eq:MMSE}
\end{equation}
This task can be approximately\footnote{The considered graph typically contains cycles which lead to loopy belief propagation that yields an approximate result.} solved utilizing message passing (belief propagation) and the sum-product algorithm \cite{yedidia2011message, loeliger2007factor,kschischang2001factor} on our factor graph. 
However, a computationally efficient method is only obtained after a series of assumptions and approximations -- described in \cite{donoho2011design} -- that yield the \ac{AMP} algorithm.
Note that \ac{AMP} performs \emph{scalar} \ac{MMSE} estimation independently for every component of $\xvec$.

\subsection{\acf{AMP}}

\ac{AMP} has been introduced in \cite{donoho2009message,maleki2010approximate,donoho2010message,donoho2011design} to efficiently solve the \ac{LASSO} problem \cite{tibshirani1996regression}, also known as basis pursuit denoising \cite{chen1998atomic}, that constitutes a non-linear convex optimization problem
\begin{equation}
\widehat{\xvec}_{\text{LASSO}}(\yvec;\lambda) = 
	\arg \min_{\widetilde{\xvec}} 
	\left\{
	\frac{1}{2}
	\left\|
	\mathbf{y} - \mathbf{A}{\widetilde{\xvec}}
	\right\|_2^2
	+
	\lambda \left\| \widetilde{\xvec} \right\|_1
	\right\}
	.
	\label{eq:lasso}
\end{equation}
The underlying intuition is to find the most accurate solution with the smallest support (motivated by the assumed sparsity of $\mathbf{x}$), where $\lambda$ allows for a trade-off between accuracy with respect to the $\ell_2$ observation error 
\mbox{$\left\| 	\mathbf{y} - \mathbf{A}\widetilde{\mathbf{x}}	\right\|_2^2$} and the sparsity\footnote{Sparsity is usually expressed by the $\ell_0$-''norm'' according to \mbox{$ \left\| \mathbf{x} \right\|_0 \leq K $}. The $\ell_1$-norm relaxation was proven in \cite{donoho2006compressed} to very often yield the same result in high dimensions, 
while introducing a favorable \emph{convex} optimization problem.} 
\mbox{$\left\| \widetilde{\mathbf{x}} \right\|_1$}
of the solution.

An illustrative approach to obtain the \ac{LASSO} from \ac{MMSE} estimator \cref{eq:MMSE} is to assume that the signal vector entries are Laplacian distributed, i.e., 
$f_{\Xscalar_\xind}(\xscalar_\xind) = f_{\text{Laplace}}(\xscalar_\xind;0,\kappa)$
with
\begin{equation}
f_{\text{Laplace}}(\xscalar; \mu, \kappa) = \frac{1}{2\kappa}\exp \left(-\frac{1}{\kappa}|\xscalar-\mu| \right)
.
\label{eq:Laplace_distribution}
\end{equation}
The zero mean Laplace distribution poses a sparsity enforcing prior, i.e., its probability mass is concentrated around zero.
Plugging the Laplace signal prior into \cref{eq:posterior_expanded} and calculating the \ac{MMSE} estimate \cref{eq:MMSE}, we obtain the \ac{LASSO} \cref{eq:lasso} with $\lambda=\sigma_\Wscalar^2/\kappa$.

In \ac{AMP}, $\lambda$ is a design parameter.
For the optimal choice of $\lambda$, it was shown in \cite{mousavi2013asymptotic} that the fixed point of the \ac{AMP} solution conicides with the \ac{LASSO} solution in the \emph{asymptotic regime} where  $\ydim/\xdim = \text{const.}$ while $\xdim,\ydim \rightarrow \infty$.

As discussed in \cite{bayati2011dynamics, montanari2012graphical}, \ac{AMP} \emph{decouples} the estimation problem associated to measurement \cref{eq:cs_measurement} into $\xdim$ uncoupled scalar problems in the asymptotic regime:
\begin{equation}
\yvec = \A \xvec + \wvec \xrightarrow{\text{asympt.}} \left\{
                \begin{array}{l}
                  u_1 = \xscalar_1 + \widetilde{\wscalar}_1 \\
									\vdots \\
									u_\xdim = \xscalar_\xdim + \widetilde{\wscalar}_\xdim
                \end{array}
\right.
,
\label{eq:decoupling}
\end{equation}
where the \emph{effective noise} asymptotically obeys 
$\widetilde{\Wscalar}_\xind \sim \mathcal{N}(0,\varc)$.
Note that it is Gaussian\footnote{This is an assumption that is satisfied in the asymptotic regime.}, and $\varc > \sigma_{\Wscalar}^2$.
Revisiting the \ac{LASSO} problem in the scalar case, i.e.,
\begin{equation}
\widehat{\xscalar}_{\text{LASSO}}(u;\lambda) = 
	\arg \min_{\widetilde{\xscalar}} 
	\left\{
	\frac{1}{2}
	\left(
	u - \widetilde{\xscalar}
	\right)^2
	+
	\lambda | \widetilde{\xscalar} |
	\right\}
	,
	\label{eq:lasso_scalar}
\end{equation}
it is known that the \emph{soft thresholding} function 
\begin{equation}
\eta(u;\tau) = 
\left\{
                \begin{array}{ll}
                  u+\tau & \text{if}\  u < -\tau \\
                  0 & \text{if}\  -\tau \leq u \leq \tau\\
                  u-\tau & \text{if}\  u > \tau
                \end{array}
\right.
\end{equation}
admits a (possibly optimal) solution to \cref{eq:lasso_scalar}, see \cite{tibshirani1996regression}.
The soft thresholding function acts as a \emph{denoiser}, i.e., it sets values below a certain threshold to zero.
This function is also found in the \ac{AMP} algorithm -- our implementation is stated by \Cref{algo:amp} -- where it is applied entry-wise on the decoupled measurements 
${\mathbf{x}}+\mathbf{A}^\transp\mathbf{r}$, i.e., on $u_\xind = \xscalar_\xind + (\A_{:,\xind})^\transp\mathbf{r}$.
The iterations are stopped once the change in the estimated signal is below a certain threshold -- controlled by $\epsilon_{\text{tol}}$ -- or the maximal number of iterations $t_\text{max}$ is reached.
For a detailed derivation of \ac{AMP}, the interested reader is refered to 
\cite{donoho2009message,maleki2010approximate,donoho2010message,donoho2011design}.

\subsection{\acf{BAMP}}

While the standard \ac{AMP} algorithm implicitly assumes a (sparsity enforcing) Laplacian signal prior, its Bayesian optimal version \ac{BAMP} allows to specify arbitrary signal priors $\pdf_{\Xscalar_\xind}(\xscalar_n)$, individually for each entry of the signal vector --- this is a key feature that will be exploited by the proposed algorithms for structured sparsity.
As before, we will stick to the main features and refer to \cite{maleki2010approximate,donoho2010message,donoho2011design} for details.

\ac{BAMP} builds on the same decoupling principle \cite{bayati2011dynamics, montanari2012graphical} as \ac{AMP}, which is valid in the asymptotic regime and approximately satisfied in finite dimensions. 
Let us discuss the decoupled scalar problem \cref{eq:decoupling},
$u_\xind = \xscalar_\xind + \widetilde{\wscalar}_\xind$,
where we know that $\widetilde{\Wscalar}_\xind \sim \mathcal{N}(0,\varc)$. 
The (entry-wise) \emph{posterior} of the decoupled problem thus reads, using Bayes' theorem,
\begin{equation}
\pdf_{\Xscalar_\xind|\Uscalar_\xind}(\xscalar_\xind|u_\xind ) =
\frac{1}{\pdf_{\Uscalar_\xind(u_\xind)}} 
\pdf_{\Uscalar_\xind|\Xscalar_\xind}(u_\xind|\xscalar_\xind)
\pdf_{\Xscalar_\xind}(\xscalar_\xind)
,
\end{equation}
where
$\pdf_{\Uscalar_\xind}(u_\xind) = 
\int_{-\infty}^{\infty} 
\pdf_{\Uscalar_\xind|\Xscalar_\xind}(u_\xind|\widetilde{\xscalar}_\xind) 
\pdf_{\Xscalar_\xind}(\widetilde{\xscalar}_\xind)
d \widetilde{\xscalar}_\xind$
and
$\pdf_{\Uscalar_\xind|\Xscalar_\xind}(u_\xind|\xscalar_\xind)=\mathcal{N}(u_\xind|\xscalar_\xind,\beta)$
.
Instead of soft thresholding, \ac{BAMP} utilizes the following functions \cite{maleki2010approximate,donoho2010message,donoho2011design}:
\begin{align}
F(u_\xind; \varc) &= \mathbb{E}_{\Xscalar_\xind}\{\Xscalar_\xind|\Uscalar_\xind=u_\xind ; \varc\} 
,
\label{eq:F}\\
G(u_\xind; \varc) &= \mathrm{Var}_{\Xscalar_\xind}\{\Xscalar_\xind|\Uscalar_\xind=u_\xind ; \varc\}
,
\label{eq:G}\\
F^{\prime}(u_\xind; \varc) &= \frac{d}{d u_\xind}F(u_\xind; \varc) 
.
\label{eq:Fprime}
\end{align}
The conditional expectation \cref{eq:F} yields the scalar \ac{MMSE} estimate of $\xscalar_\xind$ given the decoupled measurement $u_\xind$.
Our implementation of \ac{BAMP} is stated by \Cref{algo:bamp}.
Function \cref{eq:F} is applied entry-wise on a vector argument, \cref{eq:G} is typically an intermediate step to compute \cref{eq:Fprime}.
Note that for a specified signal prior, functions \cref{eq:F,eq:G,eq:Fprime} admit closed form expressions.
We will now specify those for a \emph{sparse binary} and a \emph{sparse Gaussian} signal prior, respectively.

\begin{algorithm}[t]
\caption{\acs{AMP}}
\label{algo:amp}
\begin{algorithmic}[1]
\State initialize  ${\mathbf{x}}^t=\mathbf{0}$ and $\mathbf{r}^t=\mathbf{y}$ for $t=0$
  \Do
    \State $t = t+1$ \Comment{advance iterations}
		\State $\tau = \frac{\lambda}{\sqrt{M}} \| \mathbf{r}^{t-1} \|_2$ \Comment{compute threshold}
		\State ${\mathbf{x}}^{t} = \eta ({\mathbf{x}}^{t-1}+\mathbf{A}^\transp\mathbf{r}^{t-1};\tau)$ \Comment{soft thresholding}
		\State $b=\frac{1}{M}\left\| {\mathbf{x}}^t \right\|_0$ \Comment{compute sparsity}
		\State $\mathbf{r}^t = \mathbf{y} - \mathbf{A}{\mathbf{x}}^t + b \mathbf{r}^{t-1}$ \Comment{compute residual}
  \doWhile{$\left\|\mathbf{x}^{t}-\mathbf{x}^{t-1} \right\|_2 > \epsilon_{\text{tol}} \left\|\mathbf{x}^{t-1}\right\|_2$ and $t < t_\text{max}$}
	\State return $\widehat{\mathbf{x}}=\mathbf{x}^t$ \Comment{recovered sparse vector}
\end{algorithmic}
\end{algorithm}

\begin{algorithm}[t]
\caption{\acs{BAMP}}
\label{algo:bamp}
\begin{algorithmic}[1]
\State initialize ${\mathbf{x}}^t=\mathbf{0}$ and ${\resi}^t=\mathbf{y}$  for $t=0$
  \Do
    \State $t = t+1$ 
		\State $\mathbf{u}^{t-1} = \mathbf{x}^{t-1} + \mathbf{A}^\transp \resi^{t-1}$ \Comment{decoupled measurements}
		\State $\varc^{t-1} = \frac{1}{M}\|\resi^{t-1}\|_2^2$  \Comment{effective noise estimate}
		\State $\mathbf{x}^t = F(\mathbf{u}^{t-1} ; \varc^{t-1})$   \Comment{estimate signal} 
		\State $\resi^{t} = \mathbf{y} - \mathbf{A}\mathbf{x}^t + \resi^{t-1} \frac{1}{M}\sum_{n=1}^{N} F^{\prime} (u_n^{t-1} ; \varc^{t-1})$		
  \doWhile{$\left\|\mathbf{x}^{t}-\mathbf{x}^{t-1} \right\|_2 > \epsilon_{\text{tol}} \left\|\mathbf{x}^{t-1}\right\|_2$ and $t < t_\text{max}$} 
	\State return $\widehat{\mathbf{x}}=\mathbf{x}^t$ 
\end{algorithmic}
\end{algorithm}

\begin{description}
\item[Sparse Binary Signal Prior:]

for $\xscalar_\xind \in \{0,1\}$, the prior reads
\begin{equation}
\pdf_{\Xscalar_\xind}\!(\xscalar_\xind) = \gamma_\xind \delta(\xscalar_\xind) + (1-\gamma_\xind) \delta(\xscalar_\xind-1),
\label{eq:prior_binary}
\end{equation}
where $\gamma_\xind$ indicates the probability of $\xscalar_\xind$ being a zero.
In this setting, functions \cref{eq:F,eq:G,eq:Fprime} boil down to
\begin{align}
F(u_\xind; \varc, \gamma_n) &= \frac{1}{1+\exp\left( \frac{1-2u_\xind}{2\varc} + \log\frac{\gamma_\xind}{1-\gamma_\xind} \right)}
,
\label{eq:F_sparse_binary}\\
G(u_\xind; \varc, \gamma_n) &= F(u_\xind; \varc, \gamma_n) - F(u_\xind; \varc, \gamma_n)^2
,
\label{eq:G_sparse_binary}\\
F^{\prime}(u_\xind; \varc, \gamma_n) &= \frac{1}{\varc} G(u_\xind; \varc, \gamma_n)
.
\label{eq:Fprime_sparse_binary}
\end{align}

\item[Sparse Gaussian Signal Prior:]
for $\xscalar_\xind \in \{0,\mathcal{N}(0,\varx)\}$, the prior reads
\begin{equation}
\pdf_{\Xscalar_\xind}\!(\xscalar_\xind) = \gamma_\xind \delta(\xscalar_\xind) \!+\! (1\!-\!\gamma_\xind)
\mathcal{N}(\xscalar_\xind|0,\varx),
\label{eq:prior_gaussian}
\end{equation}
where $\gamma_\xind$ indicates the probability of $\xscalar_\xind$ being a zero.
In literature, this case is also known as the \emph{Bernoulli-Gaussian} case.
Functions \cref{eq:F,eq:G,eq:Fprime} calculate as
\begin{align}
F(u_\xind; \varc, \gamma_n) &= u_\xind M(u_\xind, \gamma_\xind, q)
,
\label{eq:F_sparse_gaussian}\\
G(u_\xind; \varc, \gamma_n) &= \varc M(u_\xind, \gamma_\xind, q) + m(u_\xind, \gamma_\xind, q)
,
\label{eq:G_sparse_gaussian}\\
F^{\prime}(u_\xind; \varc, \gamma_n) &= \frac{1}{\varc} G(u_\xind; \varc, \gamma_n)
,
\label{eq:Fprime_sparse_gaussian}
\end{align}
with $q=\frac{\varx}{\varc}$ and the auxiliary functions
\begin{align}
M(u_\xind, \gamma_\xind, q) &= \frac{q}{1+q} \frac{1}{1+m(u_\xind, \gamma_\xind, q)}
,
\label{eq:M}\\
m(u_\xind, \gamma_\xind, q)  &= \frac{\gamma_\xind}{1-\gamma_\xind}\sqrt{1+q} \exp\left(-\frac{u_\xind^2}{2\varc}\frac{q}{1+q} \right)
.
\label{eq:m}
\end{align}

\end{description}
If the number of nonzero entries $K$ is known a priori, we choose
\begin{equation}
 \gamma_n = 1-\frac{K}{N}, \forall \xind \in \variables=\{1,...,\xdim\}.
\label{eq:gamma_init}
\end{equation}
In case uf unknown sparsity, one has to assume a certain sparsity and plug in an estimate for $K$.

%% file: recovery_sparse_group.tex
\begin{figure}[t]
\includegraphics[width=0.49\textwidth]{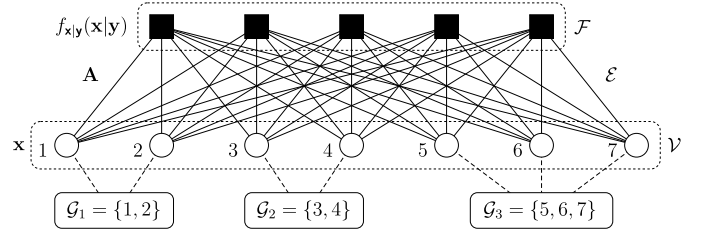} 
\caption{Factor graph with group indication.}
\label{fig:factor_graph_group}
\end{figure}

In the group sparse case, signal vector $\xvec$ is partitioned into $\numgroups$ groups such that the groups partition (non-overlapping case) the total support set 
$\variables=\{1,...,\xdim\}$:
\begin{equation}
\variables  = \bigcup_{g=1}^{\numgroups} \group_g
,
\end{equation}
where $\group_g$ contains the signal vector indices that correspond to group $g$.
In case of overlapping groups, the intersection of two different groups may contain elements.
The signal vector entries that correspond to a group are either \emph{all zero} or \emph{all nonzero} --- knowing the groups reflects the a priori knowledge of the \emph{signal structure}.
An exemplary group assignment on a factor graph is depicted in \cref{fig:factor_graph_group}.
While the groups may vary in their size $|\group_g|$, the signal vector $\xvec$ is assumed to be $K$-sparse, which typically implies that the number of nonzero groups is small.

Considering \ac{BAMP} in the sparse signal case, the priors \cref{eq:prior_binary,eq:prior_gaussian} allow to modify the probability that a signal entry $\xscalar_\xind$ is zero, for each entry of $\mathbf{x}$ individually.
This is the key feature exploited by \acf{BOSSAMP}.

\subsection{\acs{BOSSAMP} and Group Sparse Binary Signals}
\label{sec:group_sparse_binary}

Binary signals allow for a convenient computation of soft information in terms of $L$-values\footnote{$L$-values are log likelihood ratios in the context of coding. They are typically used in soft-input channel decoding or in iterative decoding.} \cite{hagenauer1995source,hagenauer1996iterative}:
\begin{equation}
L(\Xscalar_\xind) = \log \frac{P(\Xscalar_\xind=0)}{P(\Xscalar_\xind=1)} = \log \frac{\gamma_\xind}{1-\gamma_\xind} 
.
\label{eq:lval}
\end{equation}
A large positive value indicates a high probability of $\Xscalar_\xind$ being a zero, a large negative value a high probability of $\Xscalar_\xind$ being a one.
If we consider the decoupled measurements \cref{eq:decoupling}, the conditional $L$-values read, using \cref{eq:F_sparse_binary},
\begin{equation}
\begin{aligned}
L(\Xscalar_\xind|\Uscalar_\xind = u_\xind) &= \log \frac{P(\Xscalar_\xind=0|\Uscalar_\xind = u_\xind)}{P(\Xscalar_\xind=1|\Uscalar_\xind = u_\xind)}\\
&= \log \frac{1-F(u_\xind; \varc, \gamma_\xind)}{F(u_\xind; \varc, \gamma_\xind)}\\
&= \frac{1-2 u_\xind}{2\varc} + \log\frac{\gamma_\xind}{1-\gamma_\xind}
.
\label{eq:lval_cond}
\end{aligned}
\end{equation}
If we compute the conditional $L$-values in each iteration of \ac{BAMP}, i.e., 
compute the likelihood of being a zero for each signal entry estimate, we obtain for entry $\xscalar_\xind$ at iteration~$t$ (cf. line 6 of \Cref{algo:bamp}, using \cref{eq:lval_cond}):
\begin{equation}
L(\Xscalar_\xind|\Uscalar_\xind = u_\xind^{t-1}) 
= \frac{1-2 u_\xind^{t-1}}{2\varc^{t-1}} + \log\frac{\gamma_\xind^{t-1}}{1-\gamma_\xind^{t-1}}
.
\label{eq:lval_cond_bamp}
\end{equation}

The key feature of \ac{BOSSAMP} is to use the $L$-values \cref{eq:lval_cond_bamp} of iteration $t$ as \emph{extrinsic a priori information} \cite{hagenauer1996iterative} in the subsequent iteration $t+1$.
To that end, we calculate the $L$-values that accommodate the \emph{innovation} of the new iteration:
\begin{equation}
{L}_\xind^{t} =
L(\Xscalar_\xind|\Uscalar_\xind = u_\xind^{t-1}) -\underbrace{\log\frac{\gamma_\xind^{t-1}}{1-\gamma_\xind^{t-1}}}_{\overline{L}_\xind^{t-1}}
= \frac{1-2 u_\xind^{t-1}}{2\varc^{t-1}}.
\label{eq:innovation_lval_binary}
\end{equation}
To exploit the group structure, we introduce the {\bfseries binary extrinsic group update}
\begin{equation}
\begin{aligned}
\overline{L}_\xind^t 
= \groupupdate(\mathbf{u}^{t-1},\varc^{t-1},\gamma_\xind^{0})
:= 
 \overline{L}_\xind^{0} + &\sum_{l\in\group_g \backslash \xind} L_l^{t}
\\
=\log\frac{\gamma_\xind^{0}}{1-\gamma_\xind^{0}} +  &\sum_{l\in\group_g \backslash \xind} \frac{1-2 u_l^{t-1}}{2\varc^{t-1}}
,\\
\forall \xind \in \group_g
,
\forall g \in \{1,...,\numgroups\}
,
\label{eq:group_update_binary}
\end{aligned}
\end{equation}
which yields 
\mbox{$\overline{\mathbf{L}}^t = [\overline{L}_1^t,...,\overline{L}_\xdim^t]^\transp$}.
This can be interpreted as follows:
$\overline{L}_\xind^{0}$ is the static prior knowledge about the $n$-th entry, and 
$\sum_{l\in\group_g \backslash \xind} L_l^{t}$ is the extrinsic information of the rest of the group that contains the innovation of the current iteration.
If the extrinsic information provides a positive $L$-value, entry $\xscalar_\xind$ becomes more likely to be a zero rather than a one.

After the extrinsic group update, the signal prior is updated accordingly for the subsequent iteration.
We therefore introduce the {\bfseries prior update} 
\begin{equation}
\gamma_\xind^{t} 
= \priorupdate(\overline{L}_\xind^t) := \frac{1}{1+\exp\left(-\overline{L}_\xind^t\right)} 
,
\forall \xind \in \variables
,
\label{eq:prior_update}
\end{equation}
where we used $L$-value definition \cref{eq:lval}.

By including these two steps in the \ac{BAMP} algorithm, we obtain the \ac{BOSSAMP} algorithm for group sparse signals that is outlined in \Cref{algo:BOSSAMP_group_binary} 
--- functions \cref{eq:F_sparse_binary}, \cref{eq:Fprime_sparse_binary} and \cref{eq:group_update_binary} are utilized for sparse binary signals.
The zero probabilities are initialized as  $\gamma_n^0=1-\frac{K}{N},\forall n\in \variables$, according to \cref{eq:gamma_init}.

\begin{algorithm}[t]
\caption{\acs{BOSSAMP} for Group Sparse Signals}
\label{algo:BOSSAMP_group_binary}
\begin{algorithmic}[1]
\State init. ${\mathbf{x}}^t=\mathbf{0}_N$, ${\resi}^t=\mathbf{y}$ and $\gmavec^t\!=\!\mathbf{1}_{N}\!-\!\frac{K}{N}$  for $t=0$
  \Do
    \State $t = t+1$ 
		\State $\mathbf{u}^{t-1} = \mathbf{x}^{t-1} + \mathbf{A}^\transp \resi^{t-1}$ 
		\State $\varc^{t-1} = \frac{1}{M}\|\resi^{t-1}\|_2^2$ 
		\State $\mathbf{x}^t = F(\mathbf{u}^{t-1} ; \varc^{t-1}, \gmavec^{t-1})$  
		\State $\resi^{t} = \mathbf{y} - \mathbf{A}\mathbf{x}^t + \resi^{t-1} \frac{1}{M}\sum_{n=1}^{N} F^{\prime} (u_n^{t-1} ; \varc^{t-1} , \gamma_\xind^{t-1})$
		\State $\overline{\mathbf{L}}^t = \groupupdate( \mathbf{u}^{t-1} , \varc^{t-1}, \gmavec^{0} )$ \Comment{extrinsic group update}
		\State $\gmavec^{t} = \priorupdate(\overline{\mathbf{L}}^t)$ \Comment{prior update}
  \doWhile{$\left\|\mathbf{x}^{t}-\mathbf{x}^{t-1} \right\|_2 > \epsilon_{\text{tol}} \left\|\mathbf{x}^{t-1}\right\|_2$ and $t < t_\text{max}$} 
	\State return $\widehat{\mathbf{x}}=\mathbf{x}^t$ 
\end{algorithmic}
\end{algorithm}

\subsection{\acs{BOSSAMP} and Group Sparse Gaussian Signals}
\label{sec:group_sparse_gauss}

In the sparse Gaussian case, we are not able to express $L$-values directly as in the binary case.
We tackle this problem by introducing a binary latent random variable inspired by the E-step of the \ac{EM} algorithm \cite{dempster1977maximum,bishop2006pattern,xu1996convergence}.
This allows us to estimate the zero probabilities $\gamma_\xind^t$ in each iteration $t$ of \ac{BOSSAMP}.
Consider the prior distribution of the decoupled measurements $u_\xind=\xscalar_\xind+\widetilde{\wscalar}_\xind$ \cref{eq:decoupling} which can be expressed as a \emph{Gaussian mixture}\cite{bishop2006pattern,xu1996convergence}:
\begin{equation}
\begin{aligned}
\pdf_{\Uscalar_\xind}(u_\xind) 
&= 
\gamma_\xind \mathcal{N}(u_\xind|0,\varc)+(1-\gamma_\xind)\mathcal{N}(u_\xind|0,\varc+\varx)\\
&=
\sum_{i=1}^{2} \alpha_{\xind,i} 
\mathcal{N}(u_\xind|\mu_i,\sigma_i^2)
.
\label{eq:gaussian_mixture}
\end{aligned}
\end{equation}
We distinguish between two Gaussian distributions:
\begin{itemize}
\item Distribution $i=1$ is associated to the \emph{zero entries} in the original signal $\xvec$. 
The corresponding estimates $u_\xind$ solely contain the effective noise $\widetilde{\Wscalar}_\xind \sim \mathcal{N}(0,\varc)$ (i.e., $\xscalar_\xind=0$). 
We thus set \mbox{$\mu_1=0$} and \mbox{$\sigma_1^2=\varc$}.
\item Distribution $i=2$ is associated to the \emph{nonzeo entries} where $u_\xind$ contains the noisy signal entries ($\xscalar_\xind\neq0$).
Therefore, \mbox{$\mu_2=0$} and \mbox{$\sigma_2^2=\varc+\varx$}.
\end{itemize}
The mixing coefficients $\alpha_{n,i}$ determine the probability of the individual mixture components:
$\alpha_{n,1}$ is the probability of a zero entry, and $\alpha_{\xind,2}=1-\alpha_{\xind,1}$ is the probability of a nonzero entry.
In order to estimate these probabilities, a \emph{latent binary random variable} $\Latent_{\xind,i} \in \{0,1\}, i=1,2,$ is introduced:
\begin{equation}
\pdf_{\Uscalar_\xind}(u_\xind) = 
\sum_{\latent_{n,i}} \pmf_{\Latent_{\xind,i}}(\latent_{\xind,i}) \pdf_{\Uscalar_{\xind}|\Latent_{\xind,i}}(u_{\xind}|\latent_{\xind,i})
,
\label{eq:gaussian_mixture_latent}
\end{equation}
where $\Latent_{\xind,1} + \Latent_{\xind,2} = 1$ and $\pdf_{\Uscalar_{\xind}|\Latent_{\xind,i}}(u_{\xind}|\Latent_{\xind,i}=1)$ is a Gaussian distribution with mean $\mu_i$ and variance $\sigma_i^2$.
Defining the \ac{PMF} of $\Latent_{\xind,i}$ as
\begin{equation}
\pmf_{\Latent_{\xind,i}}(\latent_{\xind,i}) = (1-\alpha_{n,i}) \delta(\latent_{\xind,i}) + \alpha_{n,i} \delta(\latent_{\xind,i}-1) 
,
\end{equation}
the marginalization \cref{eq:gaussian_mixture_latent} becomes equivalent to \cref{eq:gaussian_mixture} --- 
we successfully reformulated (see \cite{bishop2006pattern}) the Gaussian mixture to involve a binary latent variable that can be estimated by the E-step of the \ac{EM} algorithm.
The E-step computes the probabilities 
\begin{equation}
\begin{aligned}
P(\Latent_{\xind,i}=1|\Uscalar_{\xind}=u_\xind) 
&=\frac{ \pmf_{\Latent_{\xind,i}}(1) \pdf_{\Uscalar_{\xind}|\Latent_{\xind,i}}(u_{\xind}|\Latent_{\xind,i}=1) }{ \pdf_{\Uscalar_\xind}(u_\xind) }\\
&=\frac{ \alpha_{n,i} \ \mathcal{N}(u_{\xind}|\mu_i,\sigma_i^2)}{\sum_{j=1}^{2} \alpha_{n,j}\  \mathcal{N}(u_{\xind}|\mu_j,\sigma_j^2) }
,
\end{aligned}
\end{equation}
which are called \emph{responsibilities}; $P(\Latent_{\xind,i}=1|\Uscalar_{\xind}=u_\xind) $ is the responsibility of mixture component $i$ for explaining observation $u_\xind$.
They are used to estimate the zero probabilities $\gamma_\xind^t$ in each iteration $t$, which is \emph{our E-step}. The estimate reads
\begin{multline}
\widetilde{\gamma}_\xind^t 
:=  P(\Latent_{\xind,1}=1|\Uscalar_{\xind}=u_\xind^{t-1} ; \alpha_{n,1}=\gamma_\xind^{t-1}) \\
=  \frac
{\gamma_\xind^{t-1} \mathcal{N}(u_\xind^{t-1}|0,\varc^{t-1})}
{\gamma_\xind^{t-1} \mathcal{N}\!(u_\xind^{t-1}|0,\!\varc^{t-1})\!+\!(1\!-\!\gamma_\xind^{t-1}) \mathcal{N}\!(u_\xind^{t-1}|0,\!\varx \!\!+\! \varc^{t-1}) }
.
\label{eq:e_step}
\end{multline}
As our latent variable $\Latent_{\xind,1}$ is binary, we can formulate the $L$-values
\begin{equation}
\begin{aligned}
L(\Latent_{\xind,1}|\Uscalar_{\xind}=u_\xind^{t-1}) 
&= \log \frac{P(\Latent_{\xind,1}=1|\Uscalar_{\xind}=u_\xind^{t-1})}{P(\Latent_{\xind,1}=0|\Uscalar_{\xind}=u_\xind^{t-1})} \\
&= \log \frac{ \widetilde{\gamma}_\xind^t  }{1- \widetilde{\gamma}_\xind^t }
\label{eq:lval_latent}
\end{aligned}
\end{equation}
that indicate how likely signal entry $\xscalar_\xind$ was to be zero (implies $\Latent_{\xind,1}=1$) given the measurement $u_\xind^{t-1}$.
Similar to \cref{eq:innovation_lval_binary}, we introduce the innovation $L$-values
\begin{equation}
\begin{aligned}
{L}_\xind^{t} &= 
L(\Latent_{\xind,1}|\Uscalar_\xind = u_\xind^{t-1}) -\log\frac{\gamma_\xind^{t-1}}{1-\gamma_\xind^{t-1}}\\
&= \log \frac{\widetilde{\gamma}_\xind^t(1-\gamma_\xind^{t-1})}{\gamma_\xind^{t-1} (1-\widetilde{\gamma}_\xind^t)} \\
&= \log \frac{ \mathcal{N}(u_\xind^{t-1}|0,\varc^{t-1}) }{  \mathcal{N}(u_\xind^{t-1}|0,\varx + \varc^{t-1}) } \\
&= \frac{1}{2}\log \frac{\varc^{t-1}+\varx}{\beta^{t-1}} - \frac{1}{2}\frac{(u_\xind^{t-1})^2\varx}{\varc^{t-1}(\varc^{t-1}+\varx)}
.
\label{eq:innovation_lval_gaussian}
\end{aligned}
\end{equation}
They are utilized for the {\bfseries Gaussian extrinsic group update}
\begin{multline}
\overline{L}_\xind^t 
= \groupupdate(\mathbf{u}^{t-1},\varc^{t-1},\gamma_{\xind}^{0})
:= 
 \overline{L}_\xind^{0} + \sum_{l\in\group_g \backslash \xind} L_l^{t}
\\
=\log\frac{\gamma_\xind^{0}}{1-\gamma_\xind^{0}} + \frac{1}{2} \!\! \sum_{l\in\group_g \backslash \xind} \!\!
\log \frac{\varc^{t-1}+\varxextr}{\varc^{t-1}} - \frac{(u_l^{t-1})^2 \varxextr}{\varc^{t-1}(\varc^{t-1}+\varxextr)}
,\\
\forall \xind \in \group_g
,
\forall g \in \{1,...,\numgroups\}
,
\label{eq:group_update_gaussian}
\end{multline}
which is similar to \cref{eq:group_update_binary}.
The subsequent prior update is the same as in the binary case, see \cref{eq:prior_update}.

The same \ac{BOSSAMP} algorithm body as stated by \Cref{algo:BOSSAMP_group_binary} is used --- note that the functions \cref{eq:F_sparse_gaussian}, \cref{eq:Fprime_sparse_gaussian} and \cref{eq:group_update_gaussian} are utilized for sparse Gaussian signals.

%% file: recovery_sparse_joint.tex
\begin{algorithm}[t]
\caption{\acs{BOSSAMP} for Jointly Sparse Signals}
\label{algo:BOSSAMP_joint_binary}
\begin{algorithmic}[1]
\State init. 
\mbox{$\xmat^t=\mathbf{0}_{\xdim \times \numblocks}$}, 
\mbox{$\gmamat^t\!=\!\mathbf{1}_{\xdim \times \numblocks}\!-\!\frac{K}{N}$}
and
\mbox{${\resi_b}^t=\yvec_b$}
\mbox{$\forall b\in \mathcal{B}=\{1,...,\numblocks\}$} and $t=0$
  \Do
	\State $t = t+1$ 
	\For{$b=1$ to $\numblocks$}  \Comment{BAMP iteration for all blocks}
		\State $\mathbf{u}_b^{t-1}= \xvec_b^{t-1} + \A^{\!(b)T} \resi_b^{t-1}$ 
		\State $\varc_b^{t-1} = \frac{1}{M}\|\resi_b^{t-1}\|_2^2$ 
		\State $\xvec_b^t = F(\mathbf{u}_b^{t-1} ; \varc_b^{t-1}, \gmavec_b^{t-1})$  
		\State $\resi_b^{t} = \yvec_b - \A^{\!(b)}\xvec_b^t + 
		\resi_b^{t-1}\! \frac{1}{M} \! \sum_{n} \! F^{\prime} (u_{n,b}^{t-1} ; \varc_b^{t-1} \! , \! \gamma_{\xind,b}^{t-1})$
	\EndFor
	\State $\overline{\mathbf{L}}^t = \groupupdate(\mathbf{U}^{t-1},\boldsymbol{\varc}^{t-1},\gmamat^{0})$ \Comment{extrinsic group update}
	\State $\gmamat^{t} = \priorupdate(\overline{\mathbf{L}}^t)$ \Comment{prior update}
  \doWhile{$\left\|\xmat(:)^{t}\!-\!\xmat(:)^{t-1} \right\|_2  \!> \! \epsilon_{\text{tol}} \left\|\xmat(:)^{t-1}\right\|_2$ and $t < t_\text{max}$} 
	\State return $\widehat{\mathbf{x}}=\mathbf{x}^t$ 
\end{algorithmic}
\end{algorithm}

In the jointly sparse case, we consider $\numblocks$ signal vectors $\xvec_b\in \mathbb{R}^\xdim, b\in \mathcal{B}=\{1,...,\numblocks\},$ that share a \emph{common support}
\begin{equation}
\support_{\xvec} \equiv \support_{\xvec_b},\forall b\in\mathcal{B}, 
\end{equation}
where $\support_{\xvec_b}$ contains the indices of the nonzero entries in $\xvec_b$.
In the most general case, the compressive measurements are formulated similar to \cref{eq:cs_measurement} as
\begin{equation}
\yvec_b = \A^{\!(b)}\xvec_b + \wvec_b
,
\label{eq:cs_measurement_block}
\end{equation}
where $\yvec_b \in \mathbb{R}^\ydim$, $\A^{\!(b)}\in \mathbb{R}^{\ydim \times \xdim}$, and $\wvec_b \in \mathbb{R}^\ydim$.
Let us collect the data blocks in matrices: 
$\ymat = [\yvec_1,...,\yvec_b,..., \yvec_\numblocks]$,
$\xmat = [\xvec_1,...,\xvec_b,..., \xvec_\numblocks]$ and
$\wmat = [\wvec_1,...,\wvec_b,..., \wvec_\numblocks]$.
If all \emph{sensing matrices are equal}, i.e., $\A \equiv \A^{\!(b)}, \forall b\in\mathcal{B}$, we can rewrite \cref{eq:cs_measurement_block} as
\begin{equation}
\ymat = \A \xmat + \wmat
.
\label{eq:cs_measurement_block_collected}
\end{equation}

\begin{figure}[t]
\includegraphics[width=0.49\textwidth]{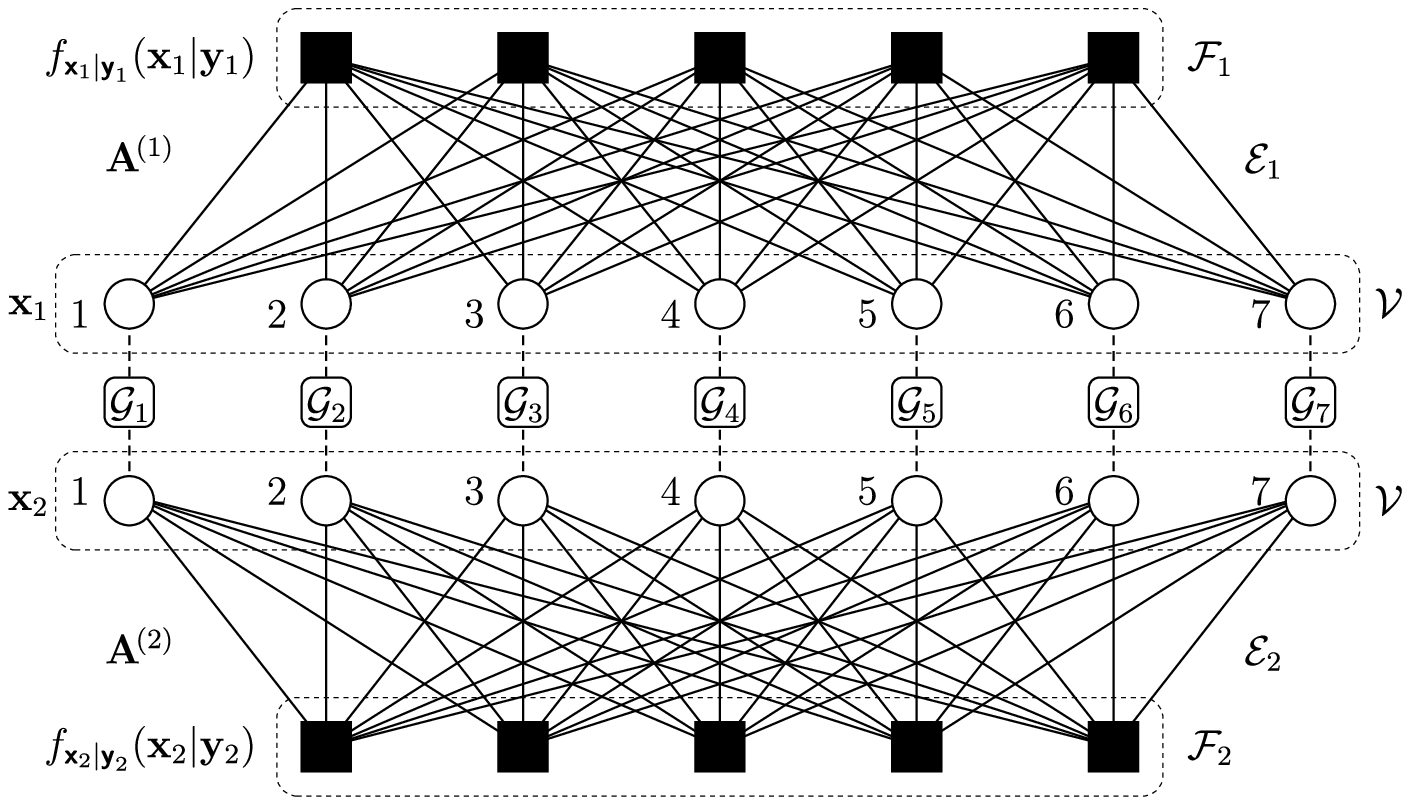} 
\caption{Factor graph of two jointly sparse signal vectors $\xvec_1$ and $\xvec_2$.}
\label{fig:factor_graph_joint}
\end{figure}

The joint sparsity is expressed by the \emph{rows} of matrix $\xmat$, whose entries are either \emph{all zero} or \emph{all nonzero}.
These rows can be interpreted as $\numgroups=\xdim$ groups, where each group $\group_g, g\in\numgroups,$ contains $|\mathcal{B}|=\numblocks$ elements.
The \ac{BOSSAMP} algorithm that exploits the joint sparsity is thus very similar to the one in the group sparse case.
An exemplary factor graph with $\numblocks=2$ blocks is depicted in \cref{fig:factor_graph_joint}.

\subsection{\acs{BOSSAMP} and Jointly Sparse Binary Signals}
\label{sec:bossamp_jointly_binary}

Compared to the group sparse case, we essentially have to extend the indexing from vectors to matrices.
In particular, $\gmamat^t=[\gmavec_1^t,...,\gmavec_\numblocks^t]$ contains the zero probabilities of iteration $t$, where $(\gmamat^t)_{\xind,b}=\gamma_{\xind,b}^t$ is the $\xind$-th entry of block $b$.
Similarly, 
$(\mathbf{U}^t)_{\xind,b} = u_{\xind,b}^t$ and
$(\overline{\mathbf{L}}^t)_{\xind,b} = \overline{L}_{\xind,b}^t$, 
and $\boldsymbol{\varc}^t = [\varc_1^t,...,\varc_\numblocks^t]^\transp$.

The regular \ac{BAMP} iteration is executed independently for each of the $\numblocks$ blocks (jointly sparse signals).
Once iteration $t$ is finished for all blocks, a 
{\bfseries collective binary extrinsic group update}, similar to \cref{eq:group_update_binary},
is executed to exploit the joint support structure among the $\numblocks$ signals:
\begin{multline}
\overline{L}_{\xind,b}^t 
= \groupupdate(\mathbf{U}^{t-1},\boldsymbol{\varc}^{t-1},\gamma_{\xind,b}^{0})
:=
\overline{L}_{\xind,b}^{0}
+
\sum_{l\in\mathcal{B} \backslash b} L_{\xind,l}^{t}
\\
=
\log \frac{\gamma_{n,b}^{0}}{1-\gamma_{n,b}^{0}}
+
\sum_{l\in\mathcal{B} \backslash b} 
\log 
\frac{1-2 u_{\xind,l}^{t-1}}{2\varc_l^{t-1}}
,\\
\forall \xind \in \variables
,
\forall b \in \mathcal{B}
.
\label{eq:joint_group_update}
\end{multline}
Afterwards, the zero probabilities are updated for the subsequent iteration by executing prior update \cref{eq:prior_update}, which is now applied entry-wise on a matrix.

The \ac{BOSSAMP} algorithm for the jointly sparse case is depicted by \Cref{algo:BOSSAMP_joint_binary}.

\subsection{\acs{BOSSAMP} and Jointly Sparse Gaussian Signals}

Considering the block structure indexing, the {\bfseries collective Gaussian extrinsic group update}, similar to \cref{eq:group_update_gaussian}, reads
\begin{multline}
\overline{L}_{\xind,b}^t 
= \groupupdate(\mathbf{U}^{t-1},\boldsymbol{\varc}^{t-1},\gamma_{\xind,b}^{0})
:=
\overline{L}_{\xind,b}^{0}
+
\sum_{l\in\mathcal{B} \backslash b} L_{\xind,l}^{t}
\\
=
\log\frac{\gamma_{\xind,b}^{0}}{1-\gamma_{\xind,b}^{0}} + \frac{1}{2} \!\! \sum_{l\in\mathcal{B} \backslash b} \!\!
\log \frac{\varc_l^{t-1}+\varxl}{\varc_l^{t-1}} - \frac{(u_{\xind,l}^{t-1})^2 \varxl}{\varc_l^{t-1}(\varc_l^{t-1}+\varxl)}
,\\
\forall \xind \in \variables
,
\forall b \in \mathcal{B}
.
\label{eq:joint_group_update_gauss}
\end{multline}
The resulting $\overline{\mathbf{L}}^t$ is then used for the subsequent entry-wise prior update \cref{eq:prior_update}.

%% file: recovery_general.tex
While the previous sections presented \ac{BOSSAMP} for the prominent examples of group/jointly sparse binary and group/jointly sparse Gaussian signals, this section will discuss the generalization to arbitrary signals.

\subsection{Arbitrary Sparse Signals}

\ac{BOSSAMP} is potentially\footnote{At this point, we can not guarantee the stability of loopy belief propagation and \ac{BOSSAMP} for all arbitrary prior distributions.} applicable to arbitrarily distributed signals ---
the crux is the utilization of the latent variable $\Latent_{\xind,i}$ as demonstrated in the Gaussian case, see \cref{sec:group_sparse_gauss}.
Soft information in terms of $L$-values is computed and exchanged extrinsically among the group (or joint) structure of the signal(s).
In the following, we consider the group sparse case; the jointly sparse case just differs in the indexing as discussed in \cref{sec:bossamp_jointly_binary}.

Consider an arbitrary sparse signal prior (similar to \cref{eq:prior_binary,eq:prior_gaussian})
\begin{equation}
\pdf_{\Xscalar_\xind}\!(\xscalar_\xind) = \gamma_\xind \delta(\xscalar_\xind) \!+\! (1\!-\!\gamma_\xind) \pdf_{a_\xind}(\xscalar_\xind)
,
\label{eq:prior_arbitrary}
\end{equation}
where $\pdf_{a_\xind}(\xscalar_\xind)$ is the distribution of the nonzero entries in $\xvec$.
Remember that random variable $\Uscalar_\xind$ is the sum of the two independent random variables $\Xscalar_\xind$ and $\widetilde{\Wscalar}_\xind$.
The prior distribution of the decoupled measurements (cf. \cref{eq:gaussian_mixture} for Gaussian case) is, therefore, obtained via convolution
\begin{equation}
\begin{aligned}
\pdf_{\Uscalar_\xind}(u_\xind) 
&= \int_{\mathbb{R}} \pdf_{\Xscalar_\xind}(v) \pdf_{\widetilde{\Wscalar}}(u_\xind-v)  dv \\
&= \gamma_\xind \mathcal{N}(u_\xind|0,\varc)+(1-\gamma_\xind)\pdf_{a_\xind\ast \widetilde{\Wscalar}_\xind}(u_\xind)
,
\label{eq:arbitrary_mixture}
\end{aligned}
\end{equation}
where $\pdf_{a_\xind\ast \widetilde{\Wscalar}_\xind}(u_\xind) = \int_{\mathbb{R}}  \pdf_{a_\xind}(v) \pdf_{\widetilde{\Wscalar}}(u_\xind-v) dv$.
Our zero probability estimates (cf. \cref{eq:e_step} for Gaussian case) now compute as
\begin{equation}
\widetilde{\gamma}_\xind^t 
=  \frac
{\gamma_\xind^{t-1} \mathcal{N}(u_\xind^{t-1}|0,\varc^{t-1})}
{\gamma_\xind^{t-1} \mathcal{N}\!(u_\xind^{t-1}|0,\!\varc^{t-1})\!+ \!(1\!-\!\gamma_\xind^{t-1}) \pdf_{a_\xind\ast \widetilde{\Wscalar}_\xind}(u_\xind^{t-1}) }
,
\label{eq:e_step_arbitrary}
\end{equation}
and the innovation $L$-values are obtained as
\begin{equation}
{L}_\xind^{t} = \log \frac{\widetilde{\gamma}_\xind^t(1-\gamma_\xind^{t-1})}{\gamma_\xind^{t-1} (1-\widetilde{\gamma}_\xind^t)} 
.
\label{eq:innovation_lval_arbitrary}
\end{equation}
The extrinsic group update is performed with these $L$-values:
\begin{multline}
\overline{L}_\xind^t 
= \groupupdate(\mathbf{u}^{t-1},\varc^{t-1},\gamma_\xind^{0})
:= 
 \overline{L}_\xind^{0} + \sum_{l\in\group_g \backslash \xind} L_l^{t}
,\\
\forall g \in \{1,...,\numgroups\}
.
\label{eq:group_update_arbitrary}
\end{multline}
Afterwards, prior update \cref{eq:prior_update} is executed.

Following this approach using, e.g., the sparse binary prior \cref{eq:prior_binary}, the resulting $L$-values \cref{eq:innovation_lval_arbitrary} coincide with \cref{eq:innovation_lval_binary}.
Note, however, that not every arbitrary distribution will entail good recovery performance in the group sparse case; the employed loopy belief propagation at the heart of \ac{BOSSAMP} may become unstable and require extensions such as damping, see \cite{heskes2002stable,heskes2004uniqueness}.

\subsection{Arbitrarily Structured Signals}

Up to now, we considered either group sparse \emph{or} jointly sparse signals.
It is straightforward to extend \ac{BOSSAMP} to be applicable to jointly sparse signals that, individually, exhibit a group structure.
To that end, the group update has to be extended as follows:
\begin{multline}
\overline{L}_{\xind,b}^t 
= \groupupdate(\mathbf{U}^{t-1},\boldsymbol{\varc}^{t-1},\gamma_{\xind,b}^{0})
:= 
 \overline{L}_{\xind,b}^{0} + \sum_{i\in\group_g \! \backslash \xind} \sum_{j\in\mathcal{B} \backslash b} L_{i,j}^{t}
,\\
\forall g \in \{1,...,\numgroups\}, \forall b \in \mathcal{B}
.
\label{eq:group_update_arbitrary_structure}
\end{multline}
As the $|\mathcal{B}|=N_B$ signals are jointly sparse, all of them exhibit the same individual group structure, i.e., the same groups $\mathcal{G}_g$. Update \cref{eq:group_update_arbitrary_structure} accounts for the group as well as the joint sparsity.

%% file: results.tex
Let us first introduce the figures of merit for comparison.
We then highlight the schemes to which we compare \ac{BOSSAMP} to.
The numerical setup is described, and the simulation results are provided, followed by a discussion.
Note that for the sake of brevity, we only present results for the \emph{group sparse} case.

\subsection{Figures of Merit}

The measurement \ac{SNR} is defined as
\begin{equation}
\text{SNR} 
= \frac{\left\| \mathbf{A}\xvec \right\|_2^2}{\mathbb{E}_{\Wvec}\left\{ \left\| \mathbf{\Wvec} \right\|_2^2 \right\}}  
= \frac{\left\| \mathbf{A}\xvec \right\|_2^2}{M \sigma_\Wscalar^2}
,
\label{eq:SNR}
\end{equation}
the noise variance $\sigma_\Wscalar^2$ is set accordingly for each realization of $\mathbf{A}$ and $\xvec$ to realize a certain \ac{SNR} during simulation.

The \ac{NMSE} between original signal $\xvec$ and its estimate (recovery) $\widehat{\xvec}$ is defined as
\begin{equation}
\NMSE = \frac{\left\| \xvec - \widehat{\xvec} \right\|_2^2}{\left\| \xvec \right\|_2^2}
,
\label{eq:NMSE}
\end{equation}
it gives indication about the overall recovery performance.

The \ac{FANMSE} is defined as
\begin{equation}
\FANMSE = \frac{\left\| \xvec_{\compsupport_\xvec} - \widehat{\xvec}_{\compsupport_\xvec} \right\|_2^2}{\left\| \xvec \right\|_2^2}
= \frac{\left\| \widehat{\xvec}_{\compsupport_\xvec} \right\|_2^2}{\left\| \xvec \right\|_2^2}
,
\label{eq:FANMSE}
\end{equation}
where the complementary signal support $\compsupport_\xvec$ contains the indices of the zero entries in $\xvec$.
The \ac{FANMSE} is a measure to quantify the strength of the false alarms in $\widehat{\xvec}$.

\subsection{Comparative Schemes}

We compare our implementations of \ac{AMP}, \ac{BAMP} and \ac{BOSSAMP} (MATLAB code will be made available at \cite{BOSSAMPcode}) to the following schemes:

\begin{description}[leftmargin=0.3cm]
\item[Group LASSO (GLASSO):] 
in order to incorporate the group structure of $\mathbf{x}$, the group \ac{LASSO} \cite{yuan2006model,friedman2010note,boyd2011distributed} replaces the $\ell_1$-norm regularization in \cref{eq:lasso} with the sum of $\ell_2$-norms of the groups:
\begin{equation}
\widehat{\xvec}_{\text{GLASSO}}(\yvec;\lambda) = 
	\arg \min_{\widetilde{\xvec}} 
	\left\{\!
	\frac{1}{2}
	\left\|
	\mathbf{y} \!-\! \mathbf{A}{\widetilde{\xvec}}
	\right\|_2^2
	+
	\lambda \!
	\sum_{g=1}^{\numgroups}
	\left\| \widetilde{\xvec}_{\group_g} \right\|_2
	\!
	\right\}
	\!
	.
	\label{eq:group_lasso}
\end{equation}
In case of $|\group_g|=1$ and $\numgroups=\xdim$, it collapses to the standard \ac{LASSO} \cref{eq:lasso}.

We use an implementation via the alternating direction method of multipliers that is described in \cite{boyd2011distributed} and whose MATLAB code is available in \cite{glasso_admm_website}.

A similar approach to solve the group \ac{LASSO} via the alternating direction method was presented in \cite{deng2013group}; however, our simulations have shown that \cite{glasso_admm_website} yields superior results.

\item[Hybrid Generalized Approx. Message Passing (HGAMP):]
\ac{GAMP} was introduced in \cite{rangan2011generalized} to extend the classical Gaussian \ac{AMP} framework -- on which we build in this paper -- to a more general setting that allows for arbitrary output channels and is thus not restricted to additive Gaussian noise in \cref{eq:cs_measurement}. 
An extension to incorporate structured sparsity was introduced in \cite{rangan2012hybrid} and is termed \ac{HGAMP}.
It was shown to outperform group orthogonal matching pursuit \cite{swirszcz2009grouped} and the group \ac{LASSO} in terms of \ac{NMSE} \cref{eq:NMSE}.
We use the MATLAB implementation of \ac{HGAMP} that is provided in \cite{gamp_website}.

\end{description}

\subsection{Numerical Setup}

For \ac{AMP}, \ac{BAMP} and \ac{BOSSAMP}, the stopping criterion was set to $\epsilon_\text{tol} = 10^{-4}$.
The maximum number of iterations was set to $t_\text{max}=100$, for all algorithms.

For \ac{AMP} \Cref{algo:amp}, we chose $\lambda=2.678 K^{-0.181}$ (\ac{NMSE} minimizing heuristic for $N=1\,000$, see \cite{mayer2014rfid}).

For {GLASSO}, the regularization parameter $\lambda$ is chosen according to the example provided in \cite{glasso_admm_website}, and the augmented Lagrangian and over-relaxation parameters were chosen as $\rho=1$ and $\alpha=1$, respectively, as suggested in \cite{glasso_admm_website}.

For \ac{HGAMP} with \emph{sparse Gaussian} signal prior, following options were selected (suggested by the toy example in \cite{gamp_website}): 
{\ttfamily step=1}, {\ttfamily removeMean=true}, {\ttfamily adaptStep=true}. 
In case of \emph{sparse binary} signal prior, following options were changed in order to mitigate numerical issues: 
{\ttfamily step=0.1}, {\ttfamily removeMean=false}.
In function {\ttfamily estim} of class {\ttfamily GrpSparseEstim.m}, the minimum and maximum value of the sparse probability {\ttfamily rho} was set to $10^{-12}$ and $1-10^{-12}$, respectively; the same values where chosen for the minimum and maxium value of {\ttfamily pr0} --- this improved the recovery performance, and mitigated numerical issues in the binary case.

For the ''{\bfseries variable SNR}'' and ''{\bfseries variable $M$}'' curves, the results are averaged over $1\,000$ random realizations.
In each realization, sensing matrix $\mathbf{A}$ and signal vector $\xvec$ are newly generated:
$\mathbf{A}$ features i.i.d. zero mean Gaussian entries with unit $\ell_2$-norm columns,
and $\xvec$ has dimension $N=1\,000$ with $K=160$ nonzero entries, entailing a zero probability of $\gamma_\xind = 0.84$.
The nonzero entries are one in the \emph{sparse binary} case \cref{eq:prior_binary}, and i.i.d. Gaussian with zero mean and variance $\varx=1$ in the \emph{sparse Gaussian} case \cref{eq:prior_gaussian}.
We consider non-overlapping equally-sized groups in $\xvec$ and compare three different cases:
\begin{itemize}
\item Group size $|\group_g| = 2, \forall g \in \{1,2,...,500\}$. With $K=160$, this implies that we have $80$ active groups out of $\numgroups=500$ total groups.
\item Group size $|\group_g| = 5, \forall g \in \{1,...,200\}$ ($32$ active groups). 
\item Group size $|\group_g| = 8, \forall g \in \{1,...,125\}$ ($20$ active groups). 
\end{itemize}

For the ''{\bfseries empirical phase transition}'' curves, we consider an undersampling $\left(\frac{M}{N}\right)$ vs. sparsity $\left(\frac{K}{M}\right)$ grid, where the values range from $0.05$ to $0.95$ with stepsize $0.05$, respectively.
At each grid point, $200$ realizations are simulated.
Let us introduce a \emph{success indicator} for each realization $r$:
\begin{equation}
S_r = 
\left\{
\begin{tabular}{cl}
1 & $\text{NMSE}_r < 10^{-4}$ \\
0 & else
\end{tabular}
\right.
.
\label{eq:success_indicator}
\end{equation}
The \emph{average success} is obtained as $\overline{S} = \frac{1}{200}\sum_{r=1}^{200}S_r$.
The empirical phase transition curves are finally obtained by plotting the $0.5$ contour of $\overline{S}$ using MATLAB function {\ttfamily contour}.

\subsection{Numerical Results and Discussion}

For convenience, \cref{tab:comparison} highlights how the various schemes utilize prior information, i.e., the sparsity, the Bayesian prior (see \cref{eq:prior_binary} and \cref{eq:prior_gaussian}), and the group structure.
In the following, we call \ac{BAMP}, \ac{BOSSAMP} and \ac{GAMP} the Bayesian message passing-based schemes.

\begin{table}
\center
\begin{tabular}{|r | c | c | c |}
\hline
 &  Sparsity & Bayesian prior & Group structure \\ \hline
{GLASSO} &  \cmark & \xmark & \cmark \\ \hline
\ac{AMP} &  \cmark & \xmark & \xmark \\ \hline
\ac{BAMP} &  \cmark & \cmark & \xmark \\ \hline
\ac{BOSSAMP} &  \cmark & \cmark & \cmark \\ \hline
\ac{HGAMP} &  \cmark & \cmark & \cmark \\ \hline
\end{tabular}
\vspace{5mm}
\caption{Utilization of prior information.}
\label{tab:comparison}
\end{table}

\begin{figure}[t]
\includegraphics[width=0.49\textwidth]{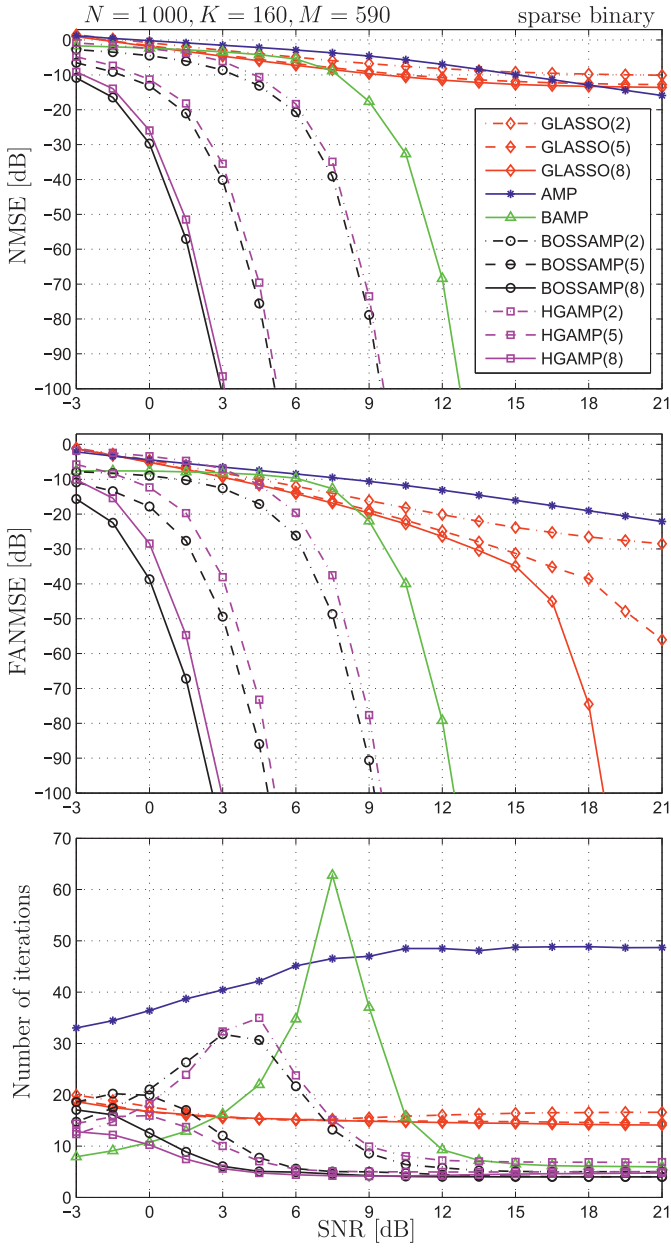} 
\caption{Variable \ac{SNR} in the sparse binary case.}
\label{fig:binary_SNR}
\end{figure}

\begin{figure}[t]
\includegraphics[width=0.49\textwidth]{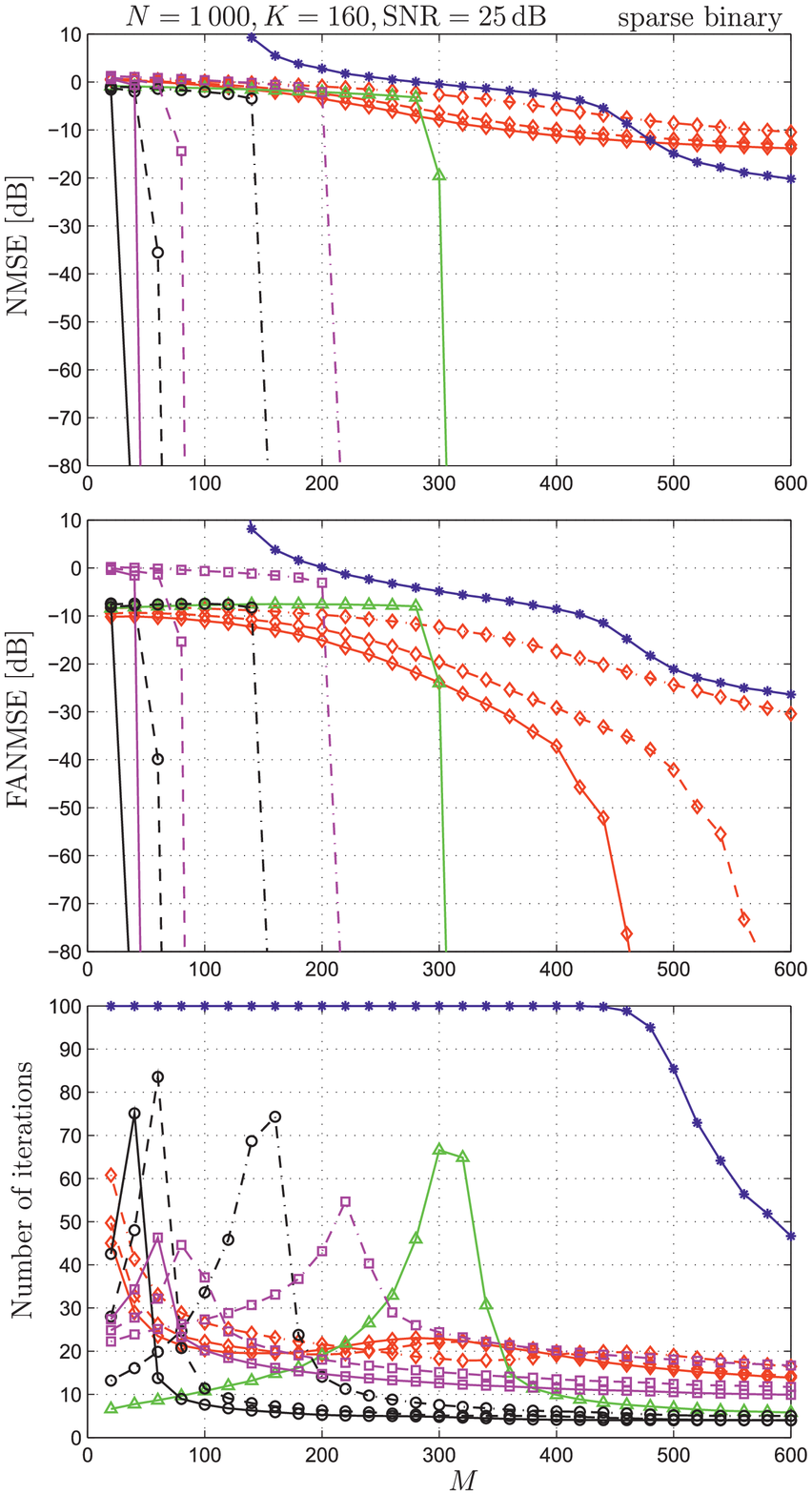} 
\caption{Variable $M$ in the sparse binary case.}
\label{fig:binary_M}
\end{figure}

\cref{fig:binary_SNR} shows the {\bfseries variable SNR} results for the \emph{sparse binary} case, where the number of measurements was fixed to $M=590$ (inspired by \cref{eq:required_measurements} with $c\approx 2$, see \cite{mayer2014rfid}).
\ac{BOSSAMP}, \ac{HGAMP} and GLASSO depend on the group size that is indicated in brackets, while \ac{AMP} and \ac{BAMP} do not exploit the group structure.
\ac{AMP} and GLASSO only include a sparsity constraint steered by $\lambda$ in \cref{eq:lasso} and \cref{eq:group_lasso}, respectively. 
\ac{BAMP}, \ac{BOSSAMP} and \ac{HGAMP} exploit the full prior knowledge, i.e., they utilize prior \cref{eq:prior_binary}.
Doing so, these schemes exhibit a steep transition to the success state ($\text{NMSE}<-40\,\text{dB}$), provided that the \ac{SNR} is sufficiently high.
It is evident that larger groups strongly improve the results of the schemes that exploit the group structure, i.e., \ac{BOSSAMP}, \ac{HGAMP} and GLASSO.
The \ac{FANMSE} draws a similar picture as the \ac{NMSE}. 
It is notable that the GLASSO is able to effectively null the false alarms, given reasonably sized groups.
Once the \ac{SNR} is large enough to ensure successful recovery, the number of iterations of the Bayesian message passing-based schemes stays very low.
The overall best recovery performance is obtained by our proposed \ac{BOSSAMP} algorithm, followed by \ac{HGAMP} and \ac{BAMP}.

\cref{fig:binary_M} shows the {\bfseries variable $M$} results for the \emph{sparse binary} case at $\text{SNR}=25\,\text{dB}$.
We observe very steep success transitions for the Bayesian message passing-based schemes.
In particular, \ac{BOSSAMP} with group size 2 yields successful recoveries above 140 measurements, while for group size 8, it only requires slightly more than 20 measurements.
In comparison, \ac{HGAMP} is successful above 200 and 40 measurements, respectively, and \ac{BAMP} requires around 300 measurements.

\begin{figure}[t]
\includegraphics[width=0.49\textwidth]{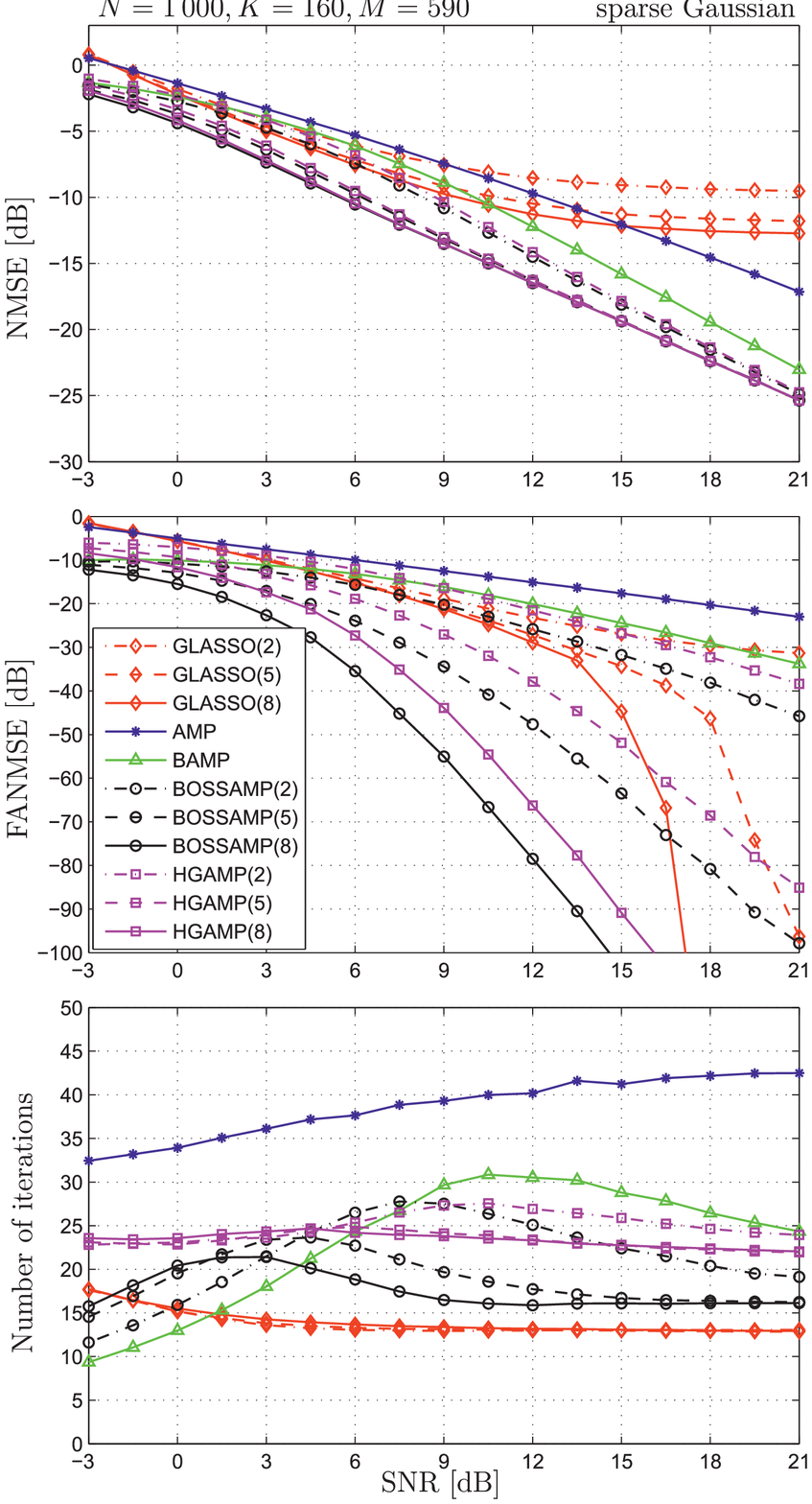} 
\caption{Variable \ac{SNR} in the sparse Gaussian case.}
\label{fig:Gaussian_SNR}
\end{figure}

\begin{figure}[t]
\includegraphics[width=0.49\textwidth]{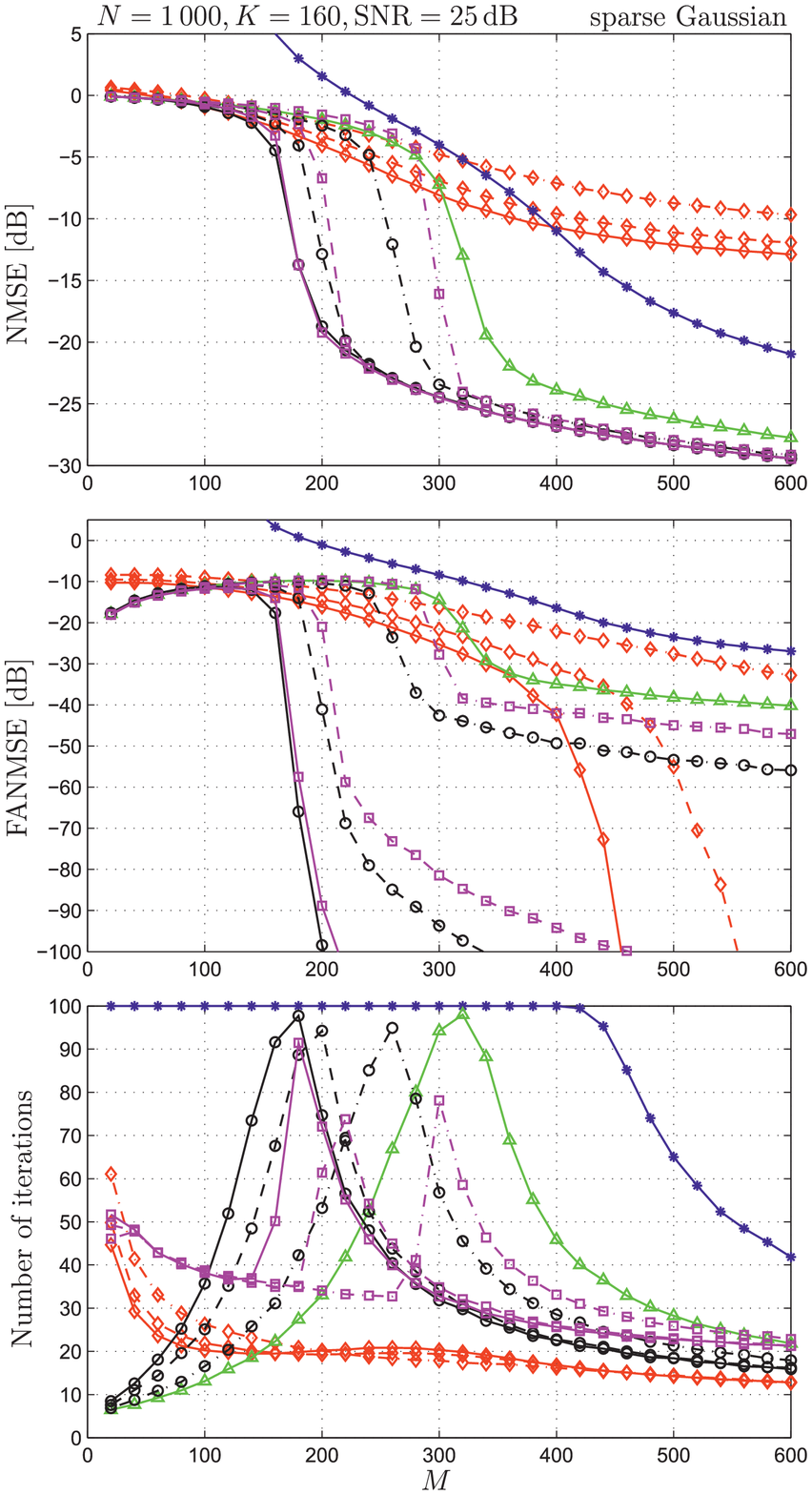} 
\caption{Variable $M$ in the sparse Gaussian case.}
\label{fig:Gaussian_M}
\end{figure}

\cref{fig:Gaussian_SNR} shows the {\bfseries variable SNR} results for the \emph{sparse Gaussian} case with $M=590$.
The message passing-based schemes do not exhibit the same steep success transitions as in the sparse binary case in \cref{fig:binary_SNR}, but a gradually decreasing \ac{NMSE} over increasing \ac{SNR}. The performance of \ac{BOSSAMP} and \ac{HGAMP} is very similar, particularly for large group sizes.
The \ac{FANMSE} shows a steeper decrease over \ac{SNR}, and it is again notable that GLASSO features a steep decay, at the expense of stagnating \ac{NMSE} (the algorithm utilizes thresholding that leads to a sparse solution but lowers the energy in the nonzero entries).
GLASSO overall requires the least number of iterations, closely followed by \ac{BOSSAMP}.
However, the Bayesian message passing-based schemes strongly outperform \ac{AMP} and GLASSO in terms of \ac{NMSE} and \ac{FANMSE}.

\cref{fig:Gaussian_M} depicts the {\bfseries variable $M$} results for the \emph{sparse Gaussian} case at $\text{SNR}=25\,\text{dB}$.
The steep success transitions of the Bayesian message passing-based schemes are back, the \ac{NMSE} is lower bounded due to finite \ac{SNR}.
It is again evident that \ac{BOSSAMP} features earlier phase transitions (requires fewer measurements) than \ac{HGAMP}, which is particularly apparent at small group sizes.
The number of iterations behave similarly for \ac{BOSSAMP}, \ac{HGAMP} and \ac{BAMP}; a distinctive peak accompanies the phase transition event, after which the iterations decrease. Due to the same message passing foundation, \ac{BOSSAMP} and \ac{HGAMP} behave similarly.

\begin{figure}[t]
\includegraphics[width=0.49\textwidth]{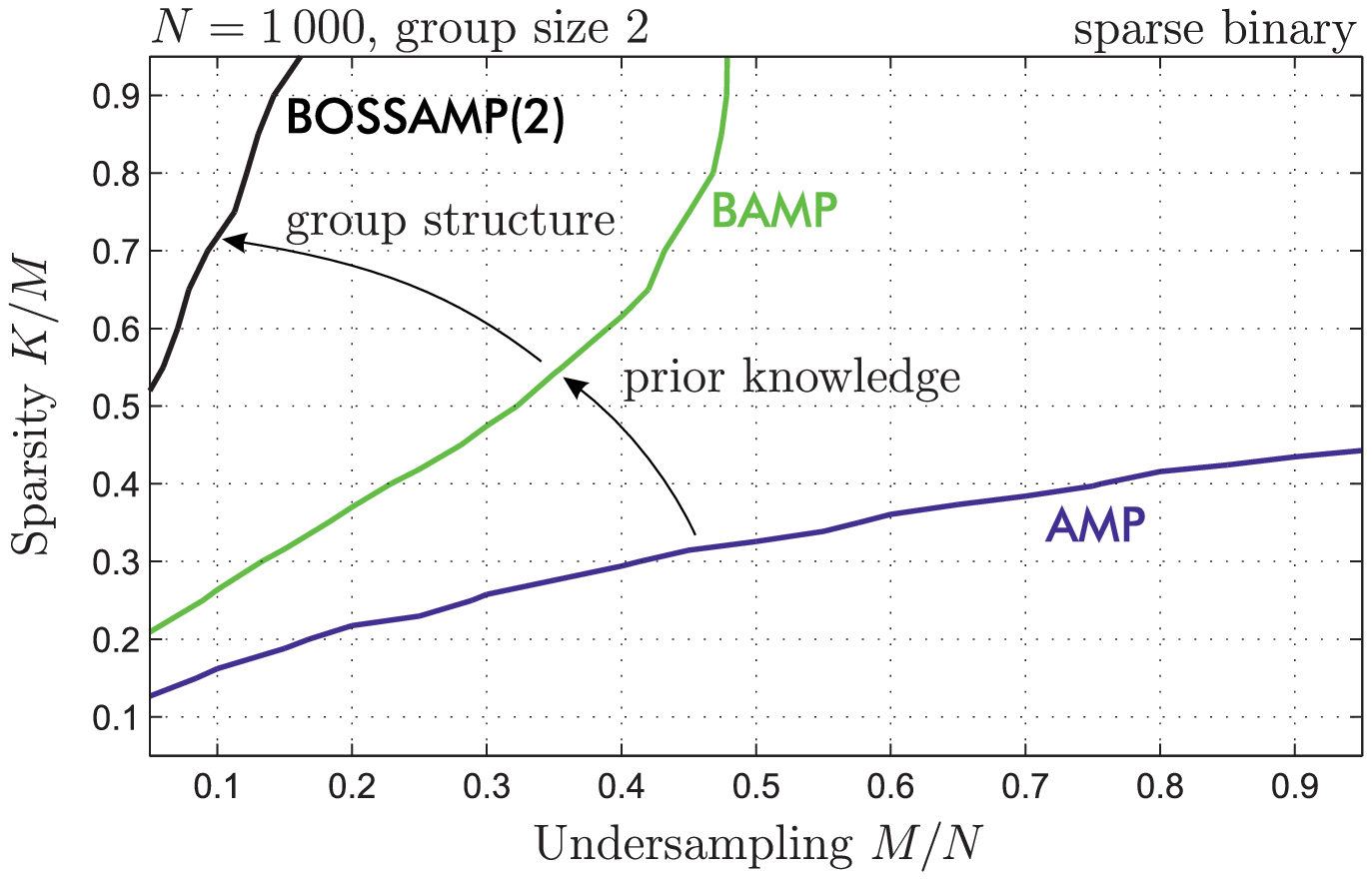} 
\caption{Empirical phase transition curves for sparse binary prior. Recoveries are successful ($\text{NMSE}<10^{-4}$) in the regime below a curve. }
\label{fig:PTC_binary}
\end{figure}

\begin{figure}[t]
\includegraphics[width=0.49\textwidth]{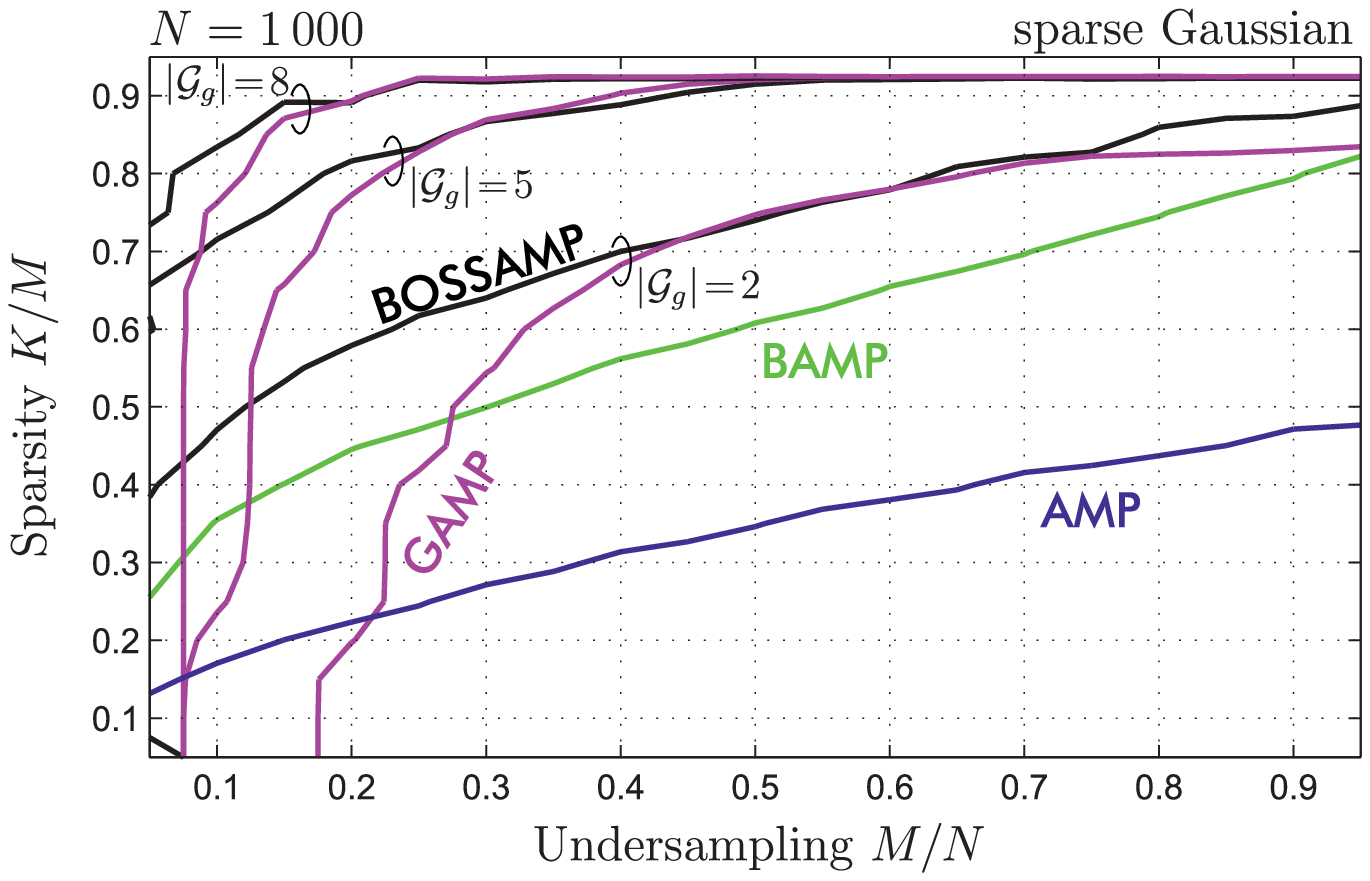} 
\caption{Empirical phase transition curves for sparse Gaussian prior. Recoveries are successful ($\text{NMSE}<10^{-4}$) in the regime below a curve. Results are plotted for three different group sizes $|\group_g|=\{2,5,8\}$.}
\label{fig:PTC_Gaussian}
\end{figure}

\cref{fig:PTC_binary} illustrates the {\bfseries empirical phase transition} curves for the \emph{sparse binary} case.
The standard \ac{AMP} algorithm exhibits the worst performance and is strongly surpassed by \ac{BAMP} that incorporates the Bayesian prior knowledge.
Additionally exploiting the group structure leads to a supreme performance which can be seen in the \ac{BOSSAMP} phase transition curve. 
Note that the group size in this example was chosen really small as $|\group_g|=2$, yet already results in a big improvement.

Finally, \cref{fig:PTC_Gaussian} illustrates the {\bfseries empirical phase transition} curves for the \emph{sparse Gaussian} case for various group sizes.
Clearly, an increase in the group size strongly improves the recovery performance of \ac{BOSSAMP} and \ac{HGAMP} and leads to earlier success transitions. 
While \ac{BOSSAMP} and \ac{HGAMP} behave similarly at large values $\frac{M}{N}$, the strongly undersampled regime causes problems for \ac{HGAMP}.

%% file: conclusion.tex
We introduced \ac{BOSSAMP}, a novel iterative algorithm to efficiently recover group sparse or jointly sparse signals.
The algorithm is based on the \ac{AMP} framework introduced by Donoho, Maleki and Montanari and exploits the known signal prior distribution.
By introducing an \emph{extrinsic group update} and a \emph{prior update} step in each iteration, the known signal structure is incorporated into the entry-wise \ac{MMSE} estimation of the standard \ac{BAMP} algorithm; considering a specific element of a group, the extrinsic group update step collects soft information from the remaining group elements.
$L$-values are accumulated according to the turbo principle and a belief about whether a specific group element was zero or nonzero arises.
According to this belief, the subsequent prior update step updates the zero probability of the prior distribution that is utilized for the \ac{MMSE} estimation in \ac{BAMP}.

We derived the group update step for the sparse binary respectively the sparse Gaussian case and provided simple closed form expressions.
Furthermore, we sketched how \ac{BOSSAMP} is potentially applicable to arbitrary sparse signals.
Simulations have shown that \ac{BOSSAMP} outperforms current state of the art algorithms, including \ac{HGAMP} that builds on the same message passing foundation.
However, \ac{HGAMP} is based on \ac{GAMP} that -- in contrast to the standard \ac{AMP} framework that we utilize in this work -- is applicable not only to the additive Gaussian noise case but to a more general class of problems, including nonlinear output relations.
While being more general, \ac{HGAMP} encompasses a more difficult implementation, and simulations suggest that due to a series of (additional) approximations, it sacrifices some performance in comparison to the standard \ac{AMP} framework \cite{maleki2010approximate,donoho2011design,donoho2010message,donoho2009message}. 

Currently, the signal prior distribution is assumed to be known. 
For the sparse Gaussian and the more general Gaussian mixture case, the parameters can be estimated using the \ac{EM}-algorithm, whose application in conjunction with \ac{GAMP} has been propagated in \cite{vila2011expectation,vila2013expectation}. 
Another promising and more general approach was proposed in \cite{guo2015near}, where Stein's unbiased risk estimator was incorporated into \ac{BAMP}.

In conclusion, we have shown that the utilization of the (known) signal structure leads to significant improvements --- on the one hand, fewer measurements are required to obtain a certain recovery performance (improved phase transition), while on the other hand, the recovery becomes more robust with respect to noise if the number of measurements is fixed.
\ac{BOSSAMP} is a versatile, easy-to-implement recovery algorithm with great performance.